\documentstyle[multicol,aps,prl,psfig]{revtex} 

\renewcommand{\narrowtext}{\begin{multicols}{2} \global\columnwidth20.5pc}
\renewcommand{\widetext}{\end{multicols} \global\columnwidth42.5pc}
\def\top#1{\vskip #1\begin{picture}(290,80)(80,500)\thinlines \put(
65,500){\line( 1, 0){255}}\put(320,500){\line( 0, 1)
{ 5}}\end{picture}}
\def\bottom#1{\vskip #1\begin{picture}(290,80)(80,500)\thinlines \put(
330,500){\line( 1, 0){255}}\put(330,500){\line( 0, -1){
5}}\end{picture}}

\def\be{\beta}
\def\ga{\gamma}
\def\de{\delta}
\def\ep{\epsilon}
\def\ve{\varepsilon}
\def\ze{\zeta}

\def\ka{\kappa}
\def\la{\lambda}

\def\si{\sigma}

\def\ch{\chi}
\def\ps{\psi}

\def\Ga{\Gamma}
\def\De{\Delta}

\def\Om{\Omega}
\def\mn{{\mu\nu}}

\def\cl{{\cal L}}

\def\prt{\partial}
\def\vev#1{\langle {#1}\rangle}
\def\expect#1{\langle{#1}\rangle}
\def\bra#1{\langle{#1}|}
\def\ket#1{|{#1}\rangle}
\def\fr#1#2{{{#1} \over {#2}}}
\def\frac#1#2{{\textstyle{{#1}\over {#2}}}}
\def\half{{\textstyle{1\over 2}}}
\def\lsim{\mathrel{\rlap{\lower4pt\hbox{\hskip1pt$\sim$}}
    \raise1pt\hbox{$<$}}}
\def\gsim{\mathrel{\rlap{\lower4pt\hbox{\hskip1pt$\sim$}}
    \raise1pt\hbox{$>$}}}
\def\sqr#1#2{{\vcenter{\vbox{\hrule height.#2pt
         \hbox{\vrule width.#2pt height#1pt \kern#1pt
         \vrule width.#2pt}
         \hrule height.#2pt}}}}

\def\hnr{\hat h}

\def\pr#1{{#1}^\prime}

\def\ol#1{\overline{#1}}

\def\a{$a_\mu$}
\def\b{$b_\mu$}
\def\c{$c_{\mu\nu}$}
\def\d{$d_{\mu\nu}$}
\def\e{$e_\mu$}
\def\f{$f_\mu$}
\def\g{$g_{\la\mu\nu}$}
\def\H{$H_{\mu\nu}$}
 
\def\lrvec#1{ \stackrel{\leftrightarrow}{#1} }

\def\kin#1{ \fr{ \expect{p^2}_{#1} }{ m^2_{#1} } }
\def\hg{ ^{199}{\rm Hg} }
\def\hgg{ ^{201}{\rm Hg} }
\def\ne{ ^{21}{\rm Ne} }
\def\xe{ ^{129}{\rm Xe} }
\def\he{ ^{3}{\rm He} }
\def\ber{ ^{9}{\rm Be}^+ }
\def\bern{ ^{9}{\rm Be} }
\def\cs{ ^{133}{\rm Cs} }
\def\li{ ^{7}{\rm Li} }

\def\tmf{ \widetilde{m}_F }
\def\hmf{ \widehat{m}_F }
\def\tmj{ \widetilde{m}_J }
\def\hmj{ \widehat{m}_J }

\def\tildeb{\tilde{b}}
\def\tildec{\tilde{c}}
\def\tilded{\tilde{d}}
\def\tildeg{\tilde{g}}

\def\x{{\hat x}} 
\def\y{{\hat y}} 
\def\z{{\hat z}}

\def\X{{\hat X}}
\def\Y{{\hat Y}}
\def\Z{{\hat Z}}

\def\kinp{ K_p }
\def\kinn{ K_n }
\def\kine{ K_e }

\newcommand{\beq}{\begin{equation}}
\newcommand{\eeq}{\end{equation}}
\newcommand{\bea}{\begin{eqnarray}}
\newcommand{\eea}{\end{eqnarray}}
\newcommand{\rf}[1]{(\ref{#1})}

\newcommand{\bdm}{\begin{displaymath}}
\newcommand{\edm}{\end{displaymath}}

\begin{document}

\title{Constraints on Lorentz Violation from Clock-Comparison Experiments}    
\author{V.\ Alan Kosteleck\'y and Charles D.\ Lane}
\address{Physics Department, Indiana University, 
          Bloomington, IN 47405, U.S.A.}
\date{IUHET 403, March 1999; accepted in Phys.\ Rev.\ D, 
scheduled for issue of Dec.\ 1 1999}

\maketitle

\vskip 0.1 truecm

\begin{abstract}
Constraints from clock-comparison experiments on violations 
of Lorentz and CPT symmetry are investigated in the context 
of a general Lorentz-violating extension of the standard model.
The experimental signals are shown to depend
on the atomic and ionic species used as clocks.
Certain experiments usually regarded as establishing comparable bounds
are in this context sensitive to different types of Lorentz violation.
Some considerations relevant to possible 
future measurements are presented. 
All these experiments are potentially sensitive 
to Lorentz-violating physics at the Planck scale.

\end{abstract}


\narrowtext

\section{Introduction}

Covariance under Lorentz transformations
is a feature of modern descriptions
of nature at the fundamental level.
These transformations include both spatial rotations and boosts,
linked through the relativistic connection between space and time.
Experimental investigations of rotation symmetry 
therefore play a crucial role in testing 
the framework of theories
such as the SU(3)$\times$SU(2)$\times$U(1) standard model
of particle physics.

Clock-comparison experiments
\cite{hughes,drever,prestage,lamoreaux,chupp,berglund,fn1}
form a class of particularly sensitive tests of rotation invariance
and hence of Lorentz symmetry.
The basic idea is to constrain possible spatial anisotropies
by bounding the variation in frequency
of a given clock as its orientation changes.
In practice,
the most precise limits are obtained by comparing the
frequencies of two different clocks as they rotate with the Earth.
The clocks used are typically atoms or ions,
and the relevant frequencies are usually those
of the light emitted or absorbed in hyperfine or Zeeman transitions.
Experiments of this type face a number of important challenges,
in particular the elimination of
systematic effects from mundane causes.
Nonetheless,
remarkable sensitivity to possible
Lorentz violations can be attained.

In the present work,
a theoretical interpretation of clock-comparison experiments 
is performed 
in the context of a general extension 
of the standard model of particle physics
incorporating a consistent microscopic theory of Lorentz violation,
including terms both even and odd under CPT
\cite{cksm}.
This standard-model extension must emerge from any underlying theory
that generates the standard model
and contains spontaneous Lorentz violation\cite{kps}.
It maintains 
both the usual gauge structure based 
on SU(3)$\times$SU(2)$\times$U(1) symmetry 
and the usual power-counting renormalizability.
It also has a variety of other desirable features,
including energy-momentum conservation,
observer Lorentz covariance,
conventional quantization,
and hermiticity,
while microcausality and positivity of the energy are expected.

{}From the perspective of the present work,
this standard-model extension is advantageous
not only because it provides 
a consistent and general theoretical framework 
for studying Lorentz violations
but more specifically because it is quantitative
and at the level of the known elementary particles.
The lagrangian of the theory is formed 
using fields for the elementary particles,
and the possible Lorentz violations for each type of particle and
interaction are controlled by parameters 
whose values are to be determined by experiment.
Since atoms and ions are composed of these elementary particles,
the behavior of different atoms and ions
under rotations and boosts
is determined by the parameters for Lorentz violation
in the theory.
It is therefore possible within this framework 
to provide a quantitative comparative analysis
of clock-comparison experiments performed with different substances
and to examine interesting possibilities for future experiments.
Both of these are undertaken in the present work.

Although many tests of Lorentz and CPT symmetry exist
\cite{pdg,cpt98,cw},
the clock-comparison ones considered here 
are among the relatively few experiments
that could be sensitive to the minuscule effects
motivating the standard-model extension.
For sensitive experiments of any type,
the standard-model extension 
provides a quantitative and coherent framework 
at the level of the standard model and 
quantum electrodynamics (QED)
within which to analyse and compare the results obtained 
and, in favorable circumstances, 
to predict possible observable signals. 
Prior to this work,
the standard-model extension has been used
to examine possible bounds on Lorentz and CPT violation
from measurements of neutral-meson oscillations
\cite{ckpv,kexpt,bexpt,ak},
from tests of QED 
in Penning traps
\cite{pennexpts,bkr,gg,hd,rm},
from photon birefringence
\cite{cfj,cksm,jk},
from hydrogen and antihydrogen spectroscopy
\cite{antih,bkr2},
and from baryogenesis
\cite{bckp}.

The structural outline of the paper is as follows.
Section \ref{theory} presents our theoretical procedures and 
discusses associated issues.
Following some general remarks,
subsection \ref{lagrangian}
is devoted to the relativistic lagrangian
and nonrelativistic hamiltonian 
used for our analysis.
The expressions for the Lorentz-violating shifts 
in atomic and ionic energy levels are 
obtained in subsection \ref{hamiltonian}.
Some comments on procedures to evaluate the
resulting expectation values are provided in
subsection \ref{coexva}.
The incorporation of geometrical effects due to the Earth's rotation
and the derivation of theoretically observable signals
is given in subsection \ref{geometry}.
Section \ref{application} 
applies this analysis,
both to published experiments
and to future possibilities.
Some comments about derivations 
relevant to specific experiments are relegated to the appendix.

\section{Theory}
\label{theory}

Clock-comparison experiments involve measurements of  
transitions between energy levels in atoms or ions.
Examining shifts in these levels is therefore 
of central interest in a theoretical analysis
of possible effects arising from Lorentz violation.
Most atoms and ions are comprised of many elementary particles
interacting together to form a system of considerable complexity,
so a complete \it ab initio \rm
calculation of energy-level shifts from the various sources of
Lorentz violation is impractical.
However,
any effects from possible Lorentz violation must be minuscule,
so theoretical calculations can proceed perturbatively
and it suffices to determine only the 
leading-order effects on the atomic or ionic energy levels. 

The Lorentz violations in the standard-model extension 
can be viewed as arising from the interaction of elementary particles 
with background expectation values of Lorentz tensor fields
in the vacuum,
somewhat like the effect of the electromagnetic field of a crystal 
on the behavior of a charged particle passing through it
\cite{cksm}. 
There are Lorentz-violating effects both in the quadratic terms
in the lagrangian and in the interactions. 
The Lorentz violations in the quadratic terms 
induce modifications to the usual free-particle propagators,
producing shifts in the conventional free-particle energies
that vary with physical properties of the particle
such as the spin and boost magnitudes and orientations.
The Lorentz violations in the interactions induce modifications to 
the vertices describing the particle interactions,
and they therefore necessarily involve the 
associated interaction coupling constant.

In the present work,
we proceed under the usual perturbative assumption that 
effects associated with free propagation are larger 
than those associated with interactions
and that the latter can therefore be disregarded 
in extracting the leading-order signals.
This approximation is likely to be good
when the elementary particles are electrons,
but may be questionable for nuclear calculations with protons or neutrons
where the strong interaction is involved.
Given this assumption,
the dominant contribution to the 
perturbative Lorentz-violating energy-level shifts 
in an atom or ion can be obtained by summing over 
individual energy shifts experienced by the component particles
as if they were freely propagating 
in the background expectation values.
The energy shifts contributed by each individual particle
can be found by taking expectation values 
of the (nonrelativistic) perturbative hamiltonian 
describing the Lorentz violation
in the multiparticle unperturbed atomic or ionic state. 

Rough dimensional estimates can be used to gain some insight 
about the relative importance 
of the perturbative approximations made.
On dimensional grounds,
the energy shift of the levels of an atom or ion 
must have the form of a product 
of some parameter for Lorentz violation 
with a function that is independent of all such parameters.
This function can be taken to be dimensionless 
(in natural units, $\hbar = c = 1$)
by absorbing a suitable power of a particle mass
in the parameter for Lorentz violation as needed.
The function can thus be approximated by 
a multivariable Taylor expansion 
in dimensionless combinations of physical quantities:
expectation values of various angular momentum operators,
relativistic correction factors involving the squared ratio 
of momentum to mass, 
and interaction energies per mass.
The expectation values of angular momenta are of order unity.
The relativistic correction factors 
are of order $10^{-2}$ for nucleons and $10^{-5}$ for electrons.
The electromagnetic-interaction energies per mass are of the order of  
$10^{-5}$ for electrons in atoms and $10^{-3}$ for protons in a 
nucleus,
while the strong-interaction energies per mass are of order $10^{-2}$.
In principle,
there is an additional dimensionless combination involving 
the ratio of the energy of the external electromagnetic field 
to the mass,
but even in magnetic fields of order 1 T
this is only of order $10^{-10}$ for electrons and $10^{-16}$
for protons.
These crude estimates suggest that 
the largest Lorentz-violation effects come from 
expectation values of angular momenta and spins.
This is confirmed by the explicit calculations that follow.

The exceptional sensitivity of clock-comparison experiments
suggests that useful bounds might in principle also be obtained 
from subleading Lorentz-violating effects,
particularly if different parameters for Lorentz violation appear.
However,
the exact calculation of subleading effects is challenging.
They arise both from relativistic corrections to the free propagation
and from corrections coupling the Lorentz violations to the interactions.
The dominant role of the strong force at the nuclear level
makes the latter corrections difficult to determine reliably.
We therefore restrict attention in the present work
to relativistic corrections arising from the free propagation
of the component particles 
in the background expectation values.
These corrections can be calculated in perturbation theory from 
subleading terms in the nonrelativistic hamiltonian.
They provide a reasonable sense of the kinds of bound
implied by subleading effects on clock-comparison experiments.

The remainder of this section 
provides the theoretical basis for our results.
Subsection \ref{lagrangian} 
presents the general quadratic relativistic lagrangian 
for a spin-$\half$ fermion,
allowing for the possibility of Lorentz violation.
It is a suitable limit of the standard-model extension
and can be used to describe 
the free propagation of the individual electrons, protons,
and neutrons forming the atom or ion of interest.
The associated nonrelativistic hamiltonian is also presented.
Subsection \ref{hamiltonian}
derives expressions for the energy-level shifts 
of a general atom or ion
by taking suitable expectation values of 
the perturbative Lorentz-violating terms in
the nonrelativistic hamiltonian.
A few more technical issues associated with 
evaluation of matrix elements in light of 
the many-body nuclear and electronic physics
are considered in subsection \ref{coexva}.
The connection to experimental observables is treated in
subsection \ref{geometry},
which examines the effect of geometrical factors 
and the Earth's rotation
on the energy-level shifts in various clock-comparison experiments.

\subsection{Lagrangian and Hamiltonian}
\label{lagrangian}

A general expression for the quadratic hermitian lagrangian 
describing a single spin-$\half$ Dirac fermion $\ps$ of mass $m$ 
in the presence of Lorentz violation is 
\cite{cksm}:
\beq
\cl = \frac{1}{2}i \ol{\ps} \Ga_\nu \lrvec{\prt^\nu} \ps 
   - \ol{\ps} M \ps
\quad , 
\label{lagr}
\eeq
where
\beq
M := m + a_\mu \ga^\mu + b_\mu \ga_5 \ga^\mu 
   + \half H_\mn \si^\mn
\quad 
\label{M}
\eeq
and
\bea
\Ga_\nu &:=& \ga_\nu + c_\mn \ga^\mu + d_\mn \ga_5 \ga^\mu 
\nonumber\\ 
&+& 
e_\nu + i f_\nu \ga_5 + \half g_{\la \mu \nu} \si^{\la \mu}
\quad .
\label{Gam}
\eea
This represents an extension of the usual lagrangian
for a massive Dirac fermion field.
The Dirac matrices 
$\left\{1, \ga_5, \ga^\mu, \ga_5\ga^\mu, \si^{\mu\nu} \right\}$
appearing in Eqs.\ \rf{M} and \rf{Gam}
all have conventional properties.

The Lorentz violation in Eq.\ \rf{lagr}
is governed by the parameters 
\a, \b, \c, \d, \e, \f, \g, and \H,
which could arise as expectation values of Lorentz tensors
following spontaneous Lorentz breaking in an underlying theory.
The hermiticity of $\cl$ means that all the parameters are real.
The parameters appearing in $M$ have dimensions of mass,
while those in $\Ga$ are dimensionless.
Both \c\ and \d\ are traceless,
while \H\ is antisymmetric
and \g\ is antisymmetric in its first two indices.
The parameters \e, \f, and \g\ are incorporated here for generality.
Gauge invariance and renormalizability exclude 
these in the standard-model extension,
so if $\ps$ represents an electron field
they are absent or suppressed relative to the others.
However, 
the situation is less clear 
if $\ps$ represents a proton or neutron
because these particles are composites of valence quarks
in a sea of other particles. 
The strong binding involved might generate effective terms
governed by appreciable parameters \e, \f, \g\
despite their absence in the standard-model extension itself.

The field operators in the terms with coefficients
\a, \b, \e, \f, and \g\ are odd under CPT,
while the others are even.
Since both the particle field
and the background tensor expectation values 
transform covariantly under rotations or boosts 
of an observer's inertial frame,
the lagrangian \rf{lagr} remains invariant
under observer Lorentz transformations.
However,
the background expectation values are unaffected 
by direct rotations or boosts of the particle or localized field
in the same observer inertial frame,
so the lagrangian transforms nontrivially
under particle Lorentz transformations
\cite{cksm}.

All these parameters are expected to be minuscule,
in which case the Lorentz-breaking effects are likely
to be detectable only in experiments of exceptional sensitivity.
Credible estimates for the order of magnitude of the parameters
are difficult to make in the absence of a realistic underlying theory.
Various sources of suppression might arise.
For example,
if the origin of the Lorentz violation lies at the Planck scale $M_P$,
one natural suppression factor would be 
some power of the ratio $r \approx m_l/M_P$,
where $m_l$ is a low-energy scale.
Another natural factor could emerge from the coupling strengths
in the underlying theory and could produce suppressions similar 
to those for the particle masses in the usual standard model,
arising from the Yukawa couplings to the Higgs scalar field.
Other substantial suppression factors might also appear.
A further potential complication is that some parameters 
might be much more heavily suppressed than others.
In what follows,
we make no specific assumptions about 
the absolute or relative magnitudes of the parameters
for Lorentz violation other than to suppose they are minuscule.

To determine the leading-order effects of the Lorentz violation,
it suffices to use a nonrelativistic
description for the particles comprising the electron cloud 
and the nuclear core
of the atoms or ions involved in the clock-comparison experiment.
We therefore need the nonrelativistic hamiltonian $h$ 
associated with the lagrangian \rf{lagr}.
The relativistic hamiltonian can be found from the lagrangian $\cl$
and the nonrelativistic momentum-space hamiltonian $h$ 
can then be derived \cite{fwpaper}
using Foldy-Wouthuysen techniques 
\cite{fw}.
The quantity of interest is the 
perturbation hamiltonian $\de h$ for Lorentz violation,
which is the difference $\de h \equiv h - \hnr$ between $h$
and the usual free-particle Foldy-Wouthuysen 
hamiltonian $\hnr$.

Including all types of operator that arise from Eq.\ \rf{lagr}
and keeping terms to second order in the Foldy-Wouthuysen expansion
for the nonrelativistic hamiltonian,
we find 
\widetext
\phantom{x}
\vskip 0.2 truecm
\bea
\de h 
&=& \left( a_0 -m c_{00} -m e_0 \right) 
+\left( -b_j + m d_{j0} - \half m \ve_{jkl}g_{kl0} 
      + \half \ve_{jkl}H_{kl} \right) \si^j  
+\left[-a_j+ m(c_{0j}+c_{j0}) +m e_j \right] \fr{p_j}{m} 
\nonumber \\ && 
+\left[ b_0 \de_{jk} - m(d_{kj} +d_{00}\de_{jk}) 
       - m \ve_{klm}(\half g_{mlj}+g_{m00}\de_{jl}) -\ve_{jkl} H_{l0} 
       \right] \fr{p_j}{m} \si^k
+\left[ m(-c_{jk}-\frac{1}{2}c_{00}\de_{jk}) \right] 
     \fr{p_j p_k}{m^2} 
  \nonumber \\
&& +\left\{
       \left[ m(d_{0j}+d_{j0})
       -\half\left( b_j+md_{j0}+\half m\ve_{jmn}g_{mn0}
       +\half\ve_{jmn}H_{mn} \right)
       \right] \de_{kl} 
       \right .
      \nonumber \\
    && 
    \qquad \qquad \qquad \qquad
       \left .
       +\half\left(b_l+\half m\ve_{lmn}g_{mn0}\right)\de_{jk}
       -m \ve_{jlm} (g_{m0k}+g_{mk0}) 
       \right\} \fr{p_j p_k}{m^2}  \si^l 
   \quad .
\label{nrham3}
\eea 
\bottom{-2.7cm}
\narrowtext
\noindent
Here,
Lorentz indices are split into timelike and spacelike 
cartesian components:
$\mu \equiv 0$ and $j = 1,2,3$.
Repeated indices are understood to be summed.
The totally antisymmetric rotation tensor $\ve_{jkl}$ satisfies 
$\ve_{123} = +1$, 
with $\ve^{jkl} = -\ve_{jkl}$ as a consequence of the
embedding of the rotation group in the Lorentz group
and the usual adoption of a metric tensor of signature $-2$.
The three-momentum of the particle is denoted by $p_j$,
while the Pauli matrices are denoted by $\si^j$
and obey $[\si^j,\si^k]=2i\ve_{jkl}\si^l$ as usual.

The leading-order terms in Eq.\ \rf{nrham3} are those unsuppressed 
by powers of $p_j/m$.
Nonrelativistic experiments with ordinary matter 
are therefore dominantly sensitive to the particular combinations
of parameters for Lorentz violation appearing in these terms.
A further separation of these combinations 
at the leading-order nonrelativistic level would require experiments 
with antimatter.
However,
the exceptional sensitivity of clock-comparison experiments
means that they could in principle place interesting (but weaker) 
bounds on further combinations of parameters for Lorentz violation
arising in subleading terms 
of the nonrelativistic hamiltonian.
Disregarding interactions,
the relevant effects arise from terms in $\de h$
of second order in $p_j/m$.
In what follows,
we investigate bounds arising from all the terms displayed
in Eq.\ \rf{nrham3}.

The nonrelativistic hamiltonian \rf{nrham3} 
describes species-specific energy shifts
depending on the spin and momentum of individual particles.
Certain other approaches to Lorentz violation
also suggest effects of this type.
Some indication of their relation to the present work
can be obtained by comparing them to 
the hamiltonian \rf{nrham3} and its associated underlying theory.
A complete review lies beyond the scope of this paper,
and we limit ourselves here to only a few remarks 
\cite{cpt98,cw}.

Among the purely phenomenological treatments
that have been widely applied to clock-comparison experiments
is the $TH\ep\mu$ formalism  
\cite{ll}.
This provides a parametrization 
of the dynamics of classical charged pointlike test particles 
in an external spherically symmetric and static gravitational field.
It has been used to probe quantitatively the foundations
of theories of gravity,
including the possibility of deviations from local Lorentz invariance.
The $TH\ep\mu$ formalism differs qualitatively in several respects from
the standard-model extension studied here.
Thus,
the latter has observer Lorentz covariance
and provides an apparently consistent theory at the quantum level
for all nongravitational forces,
but in its present formulation does not explicitly include gravity 
(although gravity is implicitly present and some aspects
of its inclusion have been investigated \cite{kps}).
Neglecting gravity,
the overlap between the theories is perhaps greatest 
in the purely electromagnetic sector,
where the $TH\ep\mu$ parameters $\mu$ and $\ep$ have similar features 
to certain components of the parameter $(k_F)_{\ka\la\mu\nu}$
in the standard-model extension
\cite{cksm}.

Several of the existing clock-comparison experiments 
have been analyzed
using models with a term of the form $K\hat{p}\cdot\vec{\si}$ 
in the hamiltonian,
where $\hat{p}$ is a unit vector in the direction of the
particle momentum with respect to some preferred frame.
A term of this type has been considered by Nielsen and Picek
\cite{np},
for example,
who regard the observed Lorentz symmetry in nature
as a low-energy manifestation
in a fundamental theory \it without \rm Lorentz invariance.
Among the terms in the hamiltonian \rf{nrham3}
are ones proportional to $\de_{jk}p_j\si^k$.
Since the standard-model extension is covariant under observer
Lorentz transformations,
its nonrelativistic hamiltonian has identical form
to lowest nonrelativistic order in all inertial frames,
and so the coefficients of these terms can emulate $K$.
Note, however, that the observer Lorentz covariance 
also ensures that the standard-model extension
strictly has no preferred frame.
There may be a frame in which certain parameters 
take a simple form
(for example, 
if $b_\mu$ is timelike 
then $b_\mu \equiv (b, \vec 0)$ in some frame),
but there is no reason \it a priori \rm 
to suppose that this frame is the same for all parameters
in the standard-model extension.

A phenomenological approach to Lorentz violation at high energies
has recently been presented by Coleman and Glashow \cite{cg}.
It assumes the existence of a preferred frame
in which there are small Lorentz-violating couplings
that are CPT and rotation invariant.
The preferred frame is identified with that of the 
cosmic microwave background,
and attention is restricted to 
renormalizable Lorentz-violating operators
that dominate at high energies.
These operators are in correspondence with 
a subset of those appearing in the standard-model extension.
For example,
with the Coleman-Glashow assumptions 
the lagrangian \rf{lagr} reduces 
in the preferred frame to one in which 
effectively only the parameters $c_{00}$ and $d_{00}$ are nonzero.

\subsection{Atomic and Ionic Energy-Level Shifts}
\label{hamiltonian}

In this subsection,
we apply the nonrelativistic hamiltonian
presented in subsection \ref{lagrangian}
to obtain perturbative shifts of atomic or ionic energy levels 
arising from Lorentz violation.

Let the atom or ion $W$ under consideration have $N_w$ particles
of type $w$, 
where $w$ is $p$ for the proton,
$n$ for the neutron, and $e$ for the electron.
The multiparticle hamiltonian describing $W$
has one (rotationally invariant) component 
arising from conventional physics
and a second (perturbative, Lorentz-violating) component $\pr{h}$
that is linear in the parameters for Lorentz violation.
The latter can be taken as the sum of the perturbative hamiltonians
for the particles comprising $W$:
\beq
\pr{h}=\sum_w\sum_{N=1}^{N_w} \de {h}_{w,N}
\quad .
\label{hprime}
\eeq
The perturbative hamiltonian $\de {h}_{w,N}$ 
for the $N$th particle of type $w$
is of the same general form as $\de h$ given in Eq.\ \rf{nrham3},
except that allowance must be made for the possibility
that the parameters for Lorentz violation depend on
the particle species $w$. 
In what follows,
this dependence is indicated by a superscript $w$
on the parameters
\a, \b, \c, \d, \e, \f, \g, and \H.

The shift of an energy level induced by the Lorentz violation
can be calculated as usual by taking the expectation value
of the perturbative hamiltonian in the appropriate 
unperturbed quantum state.
For almost all experiments of interest here,
the total angular momentum $\vec F$ of the atom or ion
and its projection along the quantization axis 
are conserved to an excellent approximation.
The quantization axis is typically determined
by the orientation of an external magnetic field,
and for simplicity we always define 
the $z$ direction in the laboratory frame
as this quantization axis.
Conservation of $\vec F$ and $F_3$ means that
the corresponding quantum numbers $F$ and $m_F$ 
can be used to label a quantum state of $W$ as $\ket{F,m_F}$,
so we proceed under this assumption.
In fact,
the rotational symmetry of one experiment of interest
\cite{prestage}
is sufficiently broken by the applied (magnetic) field
that $F$ cannot be taken as a good quantum number.
However,
in this case $\ket{F,m_F}$ can be replaced by $\ket{I,m_I}$
where $I$ is the quantum number for nuclear spin and
$m_I$ is the quantum number for its projection 
along the quantization axis.
This point is discussed further in Appendix A.

The perturbative energy shift of the state $\ket{F,m_F}$
due to Lorentz violations is given by
$\bra{F, m_F}\pr{h}\ket{F, m_F}$.
However,
only certain parts of $\pr{h}$ are relevant 
for this calculation
because the properties of $\pr{h}$ and of the states $\ket{F, m_F}$
constrain some terms to have zero expectation value.
For example,
since the relevant states of $W$ are all bound,
$\expect{\vec{p}}=0$ for all states.
More generally,
the expectation value of any odd power of momentum $\vec p$
vanishes,
and so all terms in $\pr{h}$ 
proportional to an odd power of $\vec{p}$
are irrelevant for our purposes.

Additional constraints are provided by 
the rotation properties of the states $\ket{F,m_F}$.
The expectation value of $\pr{h}$ in a state $\ket{F,m_F}$
can be written as a linear combination of terms of the form
$\bra{F,m_F} T^{(r)}_q \ket{F,m_F}$,
where $T^{(r)}_q$ represents the $q$ component of
a spherical tensor operator of rank $r$
($q=-r,\ldots,r$).
Note that individual terms in the linear combination
with $r=0$ are irrelevant to clock-comparison experiments 
because they are rotationally invariant.
The relevant terms are partially fixed by the Wigner-Eckart theorem
\cite{rose}.
This implies some terms vanish,
including any with $q\neq 0$,
and simplifies the structure of the surviving terms.
Thus,
each surviving term
is the product of two factors,
one being a ratio of Clebsch-Gordan coefficients
and the other being an expectation value
in the special state $\ket{F,m_F=F}$.
Only the former depends on $m_F$.

Restricting attention only to terms in $\pr{h}$
that generate nonzero contributions relevant 
to clock-comparison experiments,
one finds spherical tensor operators
only of rank 1 or of rank 2.
Since these operators have definite and distinct
properties under rotations,
it is useful to introduce terminology 
distinguishing their contributions to energy-level shifts.
We therefore \it define \rm
the multipolarity of an energy shift according
to the rank of the tensor from which it originates.
For example,
a dipole energy shift is one arising from
an expectation value of a tensor of rank 1,
while a quadrupole energy shift is one arising 
from an expectation value of a tensor of rank 2.
The Wigner-Eckart theorem implies 
that the energy-level shifts in $W$ can
have multipolarities at most of order $2F$.
However,
despite the generality of the theoretical framework,
no leading-order octupole or higher-order energy shifts 
can emerge from the hamiltonian $\pr{h}$
because the tensor operators involved are all of rank 2 or less.
Since monopole shifts may exist but are unobservable
in clock-comparison experiments,
only dipole and quadrupole energy shifts
are relevant to the analysis here.

Implementing the above calculations,
we find that the leading-order energy shift 
due to Lorentz violations 
of the state $\ket{F,m_F}$ 
of the atom or ion $W$ is a sum of two terms, 
\beq 
\bra{F,m_F} \pr{h} \ket{F,m_F} = 
\hmf E_d^W + \tmf E_q^W 
\quad .
\label{shift}
\eeq
In this expression,
$\hmf$ and $\tmf$
are ratios of Clebsch-Gordan coefficients
arising from the application of the Wigner-Eckart theorem and
given by
\beq
\hmf:= \fr{m_F}{F}
\quad , \qquad
\tmf:= \fr{3m_F^2-F(F+1)}{3F^2-F(F+1)}
\quad .
\label{tildemf}
\eeq
The dipole and quadrupole energy shifts
$E_d^W$ and $E_q^W$ 
are independent of $m_F$ and are given by
\bea
E_d^W &=& 
\sum_w ( 
 \be_w \tilde{b}_3^w + \de_w \tilde{d}_3^w + \ka_w \tilde{g}_d^w)
\quad , \nonumber \\
E_q^W &=& 
\sum_w ( 
 \ga_w \tildec_q^w + \la_w \tildeg_q^w )
\quad 
\label{totalEdEq}
\eea
in terms of quantities to be defined below.
The cartesian components in these and all subsequent expressions
in this subsection refer to coordinates in the laboratory frame.

In Eq.\ \rf{totalEdEq},
the various quantities with tildes 
are combinations of the parameters for Lorentz violation
appearing in the nonrelativistic hamiltonians
for the component particles $w$ of $W$.
These are the only parameter combinations
that could in principle be bounded in clock-comparison experiments
with ordinary matter.
They are defined by
\bea
\tildeb_3^w&:=& b_3^w -m_w d_{30}^w 
 + m_w g_{120}^w -H_{12}^w
 \quad,
  \nonumber \\
\tildec_q^w&:=& m_w(c_{11}^w+c_{22}^w-2c_{33}^w)
 \quad,
  \nonumber \\
\tilded_3^w&:=& m_wd_{03}^w +\half md_{30}^w-\half H_{12}^w
 \quad,
  \nonumber \\
\tildeg_d^w&:=& m_w(g_{102}^w-g_{201}^w+g_{120}^w)
 -b_3^w 
 \quad,
  \nonumber \\
\tildeg_q^w&:=& m_w(g_{101}^w+g_{202}^w-2g_{303}^w)
 \quad .
\label{bdcgtilde}
\eea
Note that each of these is chosen to have dimensions of mass.

A calculation shows that the coefficients 
$\be_w$, $\ga_w$, $\de_w$, $\ka_w$, $\la_w$
appearing in Eq.\ \rf{totalEdEq}
are linear combinations of expectation values
in the special state $\ket{F,F}$ of certain operators 
appearing in the component nonrelativistic hamiltonians
for the particles $w$ comprising $W$:
\bea
\be_w&:=&
 -\sum_{N=1}^{N_w}\langle [\si^3]_{w,N} \rangle
 \quad , 
  \nonumber \\
\ga_w&:=&
 -\fr 1 {6m_w^2}
 \sum_{N=1}^{N_w}\langle [p_1^2+p_2^2-2p_3^2]_{w,N} \rangle
 \quad , 
  \nonumber \\
\de_w&:=&
 \fr 1 {m_w^2}
 \sum_{N=1}^{N_w}\langle [p_3p_j\si^j]_{w,N} \rangle
 \quad , 
  \nonumber \\
\ka_w&:=&
 \fr 1 {2m_w^2}
 \sum_{N=1}^{N_w}\langle
 [ p_3p_j\si^j - p_j p_j \si^3 ]_{w,N} \rangle
 \quad , 
  \nonumber \\
\la_w&:=&
 \fr 1 {2m_w^2}
 \sum_{N=1}^{N_w}\langle [(p_1\si^2-p_2\si^1)p_3]_{w,N} \rangle
 \quad .
  \nonumber \\
\label{bgdkl}
\eea
The subscript $w,N$ on each operator means that it
acts on particle $N$ of type $w$.
These coefficients 
are all dimensionless.
Note that they depend on the specific atom or ion $W$.

An exact calculation of the values of the coefficients 
$\be_w$, $\ga_w$, $\de_w$, $\ka_w$, $\la_w$
is typically infeasible,
in part due to the determining role played by the nuclear forces.
Some comments about evaluating these coefficients
can be found in subsection \ref{coexva}.
On dimensional grounds
a nonzero value of $\be_w$ is likely to be of order unity,
while nonzero values of the other quantities 
are suppressed by a factor 
$K_w := {\expect{p^2}_w}/{m_w^2}$,
roughly given by $K_p\approx K_n\simeq 10^{-2}$
and $K_e\simeq 10^{-5}$.

\subsection{Comments on Expectation Values}
\label{coexva}

In this subsection,
some aspects of the evaluation of the coefficients 
$\be_w$, $\ga_w$, $\de_w$, $\ka_w$, $\la_w$
defined in Eq.\ \rf{bgdkl} are considered.
Although exact results cannot typically be derived,
partly because no exact treatment of nuclear forces is available,
some statements based on symmetry arguments
can be made despite the absence of
precise knowledge of the electronic, nuclear,
atomic, or ionic wave functions.
For some special cases and within certain approximations,
explicit results for the angular dependences of
the coefficients in Eq.\ \rf{bgdkl} can be obtained.
Under suitable circumstances,
some of the coefficients can be shown to vanish 
or to be independent of one or more of the particle species $w$.

Consider first the special case of an atom $W$
in which the electrons form a closed shell.
To a good approximation,
the expectation values in $\ket{F, F}$
appearing in Eq.\ \rf{bgdkl}
can then be replaced by expectation values 
in the state $\ket{I,I}$,
where $I$ is the quantum number
for the nuclear spin.
Following the discussion in the previous subsection,
the maximal multipolarity of the energy shifts is $2I$
and only dipole and quadrupole energy shifts are observable.
Thus,
any nucleus with $I=0$ has no observable effects.
A nucleus with $I=\half$ may have 
nonzero dipole energy shifts $E_d^W$,
but $E_q^W$ must vanish.
All other nuclei may have both dipole and quadrupole shifts. 

Further considerations based on Eq.\ \rf{bgdkl} 
are needed to determine the specific dependence of the shifts 
on the proton and neutron parameters for Lorentz violation.
One possibility is to work within a nuclear shell model
\cite{bm,mayer,nucshell}.
Consider the special case where
$W$ has a closed electronic shell,
and where a single valence nucleon of one species lies
outside closed proton and neutron shells. 
To a good approximation,
the expectation values in $\ket{F, F}$
appearing in Eq.\ \rf{bgdkl}
can then be replaced by expectation values 
in the one-nucleon state $\ket{j,j}$,
where $j = l \pm \half$ is the total angular momentum 
of the valence nucleon $w$ and $l$ is the quantum number
for its orbital angular momentum.
This implies that the values 
of the coefficients in Eq.\ \rf{bgdkl}
can be nonzero only for this nucleon.
After some calculation,
we find for $j = l+\half$ the result
$$
\be_w = -1 , ~ 
\ga_w = -\fr 1 3 \fr{l}{(2l+3)}\kin{w} , ~~ 
\de_w = \fr{1}{(2l+3)} \kin{w} , 
$$
\beq
\ka_w = -\fr{(l+1)}{(2l+3)}\kin{w} ,
\quad  \qquad
\la_w = 0 
\quad ,
\label{singlebep}
\eeq
while for $j=l-\half$ we find
$$
\be_w = \fr{(2l-1)}{(2l+1)} , \quad 
\ga_w = -\fr 1 3 \fr{(l-1)}{(2l+1)}\kin{w} , 
$$
$$
\de_w = -\fr{3(2l-1)}{(2l+1)(2l+3)} \kin{w} ,
$$
\beq
\ka_w = \fr{l(2l-1)}{(2l+1)(2l+3)}\kin{w} , 
\quad
\la_w = 0 .
\label{singlebm}
\eeq
In these expressions,
the expectation value $\expect{p^2}_w$ 
is in the radial wave function.

Equations \rf{singlebep} and \rf{singlebm} hold
in the general case when the electronic shell is closed
and the nucleus can be described by the Schmidt model
\cite{schmidt,blatt}.
In this model,
a single nucleon is assumed to carry the entire
angular momentum of the nucleus.
In the above equations, 
$j$ then becomes the nuclear spin $I$ 
and $l$ becomes the quantum number for the
orbital angular momentum assigned to the single Schmidt nucleon.
The above equations also apply to the electronic structure 
of an atom or ion in the special case where a single valence electron 
of orbital angular momentum $l$ and total angular momentum $j$
lies outside a closed shell.

More complex models can be used to gain further insight.
As an explicit example,
we consider $\li$,
which was used in both of the original clock-comparison experiments
\cite{hughes,drever}.
An approximate wave function for the $\li$ nucleus can be found
\cite{rb}
using a model 
in which two of the protons and two of the neutrons combine to
form an alpha-particle core, 
leaving a single valence proton and two valence neutrons.
The nuclear ground state has spin $I=\fr{3}{2}$,
so nonzero dipole and quadrupole energy shifts
$E_d^{\li}$, $E_q^{\li}$ are both possible in principle.
Within the model,
an approximation to the wave function of the nucleus is 
\beq
\ps^{\li} = C_1 (^1D,{^2P}) + C_2 (^1 S,{^2P})
\quad,
\label{liwave function}
\eeq
where $C_1\simeq 0.681$ and $C_2\simeq 0.732$ are constants.
Each term in parentheses represents 
a multiparticle component wave function
labeled as $ ({^{2 S_n +1}L_n} , {^{2 S_p +1}L_p)} $,
where $S_p$, $S_n$ are total spins 
and $L_p$, $L_n$ are total orbital angular momenta 
for the valence proton and neutrons.

This wave function can be used to calculate explicitly
the coefficients appearing in Eq.\ \rf{bgdkl},
but the result provides relatively little insight. 
It is of more direct interest to note that
the wave function \rf{liwave function}
indicates that $S_n=0$ and $S_p=\half$.
All the operators whose expectation values
produce the dipole shift $E_d^W$ in Eq.\ \rf{totalEdEq}
involve spin.
It therefore follows within this model that $E_d^{\li}$ 
is independent of the neutron parameters 
for Lorentz violation but does depend on proton ones.
However,
the quadrupole shift $E_q^W$ in Eq.\ \rf{totalEdEq}
involves the purely spatial operators appearing in 
the definition of $\ga_w$ in Eq.\ \rf{bgdkl}.
According to the wave function \rf{liwave function},
this is expected to produce a nonzero contribution for both 
$\ga_n^{\li}$ and $\ga_p^{\li}$
because terms with $L_n=2$ and $L_p=1$ appear.

This calculation can also be used to illustrate the dangers
of relying on a particular model to deduce details
of the origin of possible dipole or quadrupole shifts.
A further refinement of the $\li$ nuclear wave function
\cite{rb}
produces an additional term $C_3 ( ^3P,^2P )$,
with $C_3 \simeq 0.1$.
The extra term has $S_n=1$,
indicating that $E_d^{\li}$ {\it does} depend 
on neutron parameters,
although in a partially suppressed way.
This calculation also shows that 
care is required in applying results
from a simple nuclear shell model.
The ground-state properties of any odd-mass nucleus $W$ 
with an even number of neutrons 
are supposed to be determined entirely by the protons,
which would imply that both $E_d^W$ and $E_q^W$ are independent
of neutron parameters.
However,
this is not strictly correct.
A counterexample is provided by $\li$,
as above.
A similar issue arises for the ground-state properties 
of an odd-mass nucleus with an even proton number,
supposedly determined entirely by the neutrons.
A counterexample here 
is provided by the $\bern$ nucleus:
using a multiparticle wave function
\cite{rb},
a calculation shows that $E_q^{\bern}$ 
does in fact depend on proton parameters.

Despite the obstacles to definitive calculations 
of the coefficients in Eq.\ \rf{bgdkl},
some results holding under relatively mild assumptions
can be obtained. 
For example,
the Wigner-Eckart theorem can be used to show that
closed shells of particles make no contributions
to either $E_d^W$ or $E_q^W$.
A closed shell for some angular momentum $J$
has all substates $\ket{J, m_J}$ occupied,
so the contribution $\De E_{J,r,q}$
from a closed shell to the energy shift caused by 
a spherical tensor operator $T^{(r)}_q$
of rank $r$ 
($q=-r,\ldots,r$)
is given by
\beq
\De E_{J,r,q} = 
\sum_{m_J=-J}^{J} \bra{J,m_J} T^{(r)}_q \ket{J,m_J}
\quad .
\label{totalenergy}
\eeq
By the Wigner-Eckart theorem,
we find
\beq
\De E_{J,r,q} = \de_{q0} \bra{J,J} T^{(r)}_0 \ket{J,J}  
     \sum_{m_J=-J}^{J} \hat c_{Jm_Jr0}
\quad .
\label{totalenergy2}
\eeq
The coefficients
$\hat c_{Jm_Jr0}$
are ratios of Clebsch-Gordan coefficients.
For the cases $r= 1,2$ of interest
we find $\hat c_{Jm_J10}= \hmj$ and $\hat c_{Jm_J20}= \tmj$,
where $\hmj$ and $\tmj$ are given in Eq.\ \rf{tildemf}.
Explicit evaluation of the sum in Eq.\ \rf{totalenergy}
for these two cases then gives the claimed result,
$\De E_{J,1,q} = \De E_{J,2,q} = 0$. 

More general cases,
where $W$ has nontrivial electronic structure
and contributions from multiple nucleons,
could also be analyzed using the approaches in this subsection 
whenever a decomposition of the wave function $\ket{F,F}$
into a sum of multiparticle product wave functions
provides an adequate description of the atom or ion.
It then follows that
the angular dependences of the quantities defined
in Eq.\ \rf{bgdkl} can in principle be calculated
in terms of Clebsch-Gordan coefficients
and the quantum numbers for the orbital and spin angular momenta
of the component fermions of $W$.

\subsection{Geometry and Time Dependence}
\label{geometry}

The components of the parameters for Lorentz violation
appearing in Eqs.\ \rf{totalEdEq} and \rf{bdcgtilde}
are defined in the laboratory frame.
Since this frame rotates with the Earth,
the components vary in time $t$ with a periodicity 
that depends on the Earth's sidereal rotation frequency 
$\Om \simeq{2\pi}$/(23 h 56 min).  
Clock-comparison experiments typically bound the amplitude 
of the time variation of a transition frequency,
which here is related to a difference between energy shifts 
of the form $\bra{F,m_F} \pr{h} \ket{F,m_F}$.
Next,
we determine the time dependence of the energy levels
in terms of the parameters for Lorentz violation.

The first step is to introduce suitable
bases of vectors for a nonrotating frame
and for the laboratory frame.
In what follows,
the basis in the nonrotating frame is denoted
$(\X,\Y,\Z)$,
while that in the laboratory frame is denoted 
$(\x,\y,\z)$.

For the nonrotating frame,
the rotation axis of the Earth
provides a natural choice of $\Z$ axis.
Astronomers define celestial equatorial coordinates
\cite{celestial}
called declination and right ascension,
which we use to fix the $\X$ and $\Y$ axes.
The $\Z$ axis corresponds to declination 90$^\circ$.
We define $\X$ to have both declination and right ascension 0$^\circ$,
while $\Y$ has declination 0$^\circ$ and 
right ascension $90^\circ$.
Then,
$(\X,\Y,\Z)$ forms a right-handed orthonormal basis,
with the basis vectors $\X$ and $\Y$ lying in the plane
of the Earth's equator.
To the extent that 
precession of the Earth's axis can be neglected
\cite{fn2}, 
this basis is constant in time.
It is also independent of any 
particular clock-comparison experiment.

For the laboratory frame,
we take a natural definition of the $\z$ axis 
as the quantization axis of the atoms or ions involved
in the specific experiment in question.
This direction typically differs for different experiments,
so the basis $(\x,\y,\z)$ does too.
The basis $(\x,\y,\z)$ also varies in time,
and the vector $\z$ precesses about $\Z$ 
with the Earth's sidereal frequency $\Om$.
A nonzero signal in a clock-comparison experiment
preferentially requires that $\z$ \it not \rm be parallel to $\Z$,
since otherwise the time variation of the signal
arises only from the precession of the Earth's axis
and is heavily suppressed.
In what follows,
we therefore assume the angle $\ch\in(0,\pi)$
given by $\cos{\ch}=\z\cdot\Z$ is nonzero.
We choose time $t=0$ such that $\z(t=0)$
lies in the first quadrant of the $\X$-$\Z$ plane,
and we define $\x$ to be perpendicular to $\z$ 
and to lie in the plane spanned by $\z$ and $\Z$:
$\x:=\z \cot\ch - \Z\csc\ch$.
Then, 
a right-handed orthonormal basis is obtained with the definition
$\y:=\z\times\x$.
With these choices,
the $\y$ axis always lies in the plane of the Earth's equator
and is thus perpendicular to $\Z$. 
Since the laboratory frame rotates about the $\Z$ axis
with frequency $\Om$,
$\y$ coincides with $\Y$ once every (sidereal) day.

The two sets of basis vectors are shown in Fig.\ 1.
To ease visualization,
the basis $(\x,\y,\z)$ has been translated from 
the surface of the globe to the center, 
so the origins of the two basis sets coincide.
The rotation of the Earth is nonrelativistic
to a good approximation,
since a point on the Earth's equator moves 
with respect to the rotation axis
at about $10^{-6}$ lightspeed.
For most purposes
the associated relativistic effects can therefore be ignored,
and a nonrelativistic transformation 
between the two bases suffices. 
It is given by
\beq
\left( 
  \begin{array}{c}
    \x \\ \y \\ \z
  \end{array}
\right)
=\left( 
  \begin{array}{ccc}
    \cos{\ch}\cos{\Om t} & \cos{\ch}\sin{\Om t} & -\sin{\ch} \\
    -\sin{\Om t} & \cos{\Om t} & 0 \\
    \sin{\ch}\cos{\Om t} & \sin{\ch}\sin{\Om t} & \cos{\ch}
  \end{array}
\right)
\left(
  \begin{array}{c}
    \X \\ \Y \\ \Z
  \end{array}
\right)
\label{coordmatrix}
\eeq
with the above basis definitions.
This transformation can be used directly 
to obtain the time variation of the parameters for Lorentz violation.

\begin{figure}
\centerline{\psfig{figure=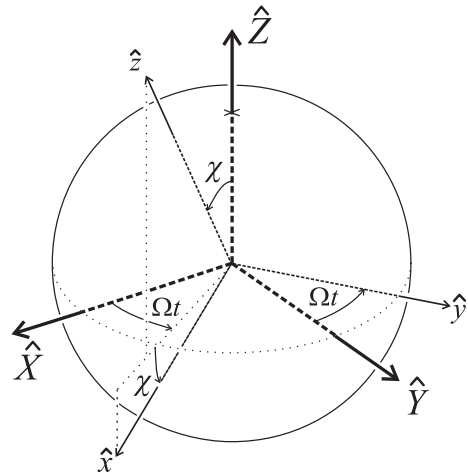,width=0.7\hsize}}
\caption{Transformation of coordinates.}
\label{fig:Figure1}
\end{figure}

To express the results in a relatively compact form,
it is convenient to introduce nonrotating-frame analogues
of the parameters in Eq.\ \rf{bdcgtilde}.
We define
\bea
\tildeb_J&:=& b_J -m d_{J0} 
 +\half m\ep_{JKL} g_{KL0} -\half \ep_{JKL} H_{KL}
 \quad,
  \nonumber \\
\tildec_{Q}&:=& m(c_{XX}+c_{YY}-2c_{ZZ})
 \quad,
  \nonumber \\
\tildec_{Q,J} &:=& m(c_{JZ}+c_{ZJ}) 
\quad , \qquad J=X,Y 
 \quad,
  \nonumber \\
\tildec_-&:=& m(c_{XX}-c_{YY}), \quad
\tildec_{XY}:= m(c_{XY}+c_{YX})
 \quad,
  \nonumber \\
\tilded_J&:=& m(d_{0J}+d_{J0})
 -\half(md_{J0}+\half\ve_{JKL}H_{KL})
 \quad,
  \nonumber \\
\tildeg_{D,J} &:=& m \ve_{JKL} (g_{K0L}+\half g_{KL0})
 -b_J 
 \quad,
  \nonumber \\
\tildeg_{Q}&:=& m(g_{X0X}+g_{Y0Y}-2g_{Z0Z})
 \quad,
  \nonumber \\
\tildeg_{Q,J} &:=& m(g_{J0Z}+g_{Z0J}) 
\quad , \qquad J=X,Y 
 \quad,
  \nonumber \\
\tildeg_-&:=& m(g_{X0X}-g_{Y0Y}), \quad 
\tildeg_{XY}:=  m(g_{X0Y}+g_{Y0X}) .
 \nonumber \\
\label{bdcgtildenonrot}
\eea
Here,
spatial indices in the nonrotating frame 
are denoted by $J = X, Y, Z$ except where indicated,
the time index is denoted $0$,
and $\ep_{JKL}$ is the nonrotating-frame analogue of $\ep_{jkl}$
obeying $\ep_{XYZ}=+1$.
The label $w$ is suppressed for simplicity.

With these definitions,
the transformation matrix in Eq.\ \rf{coordmatrix}
can be used to express the
time dependence of the parameters for Lorentz violation
in the laboratory frame:
\bea
\tilde{b}_3 &=& \tilde{b}_Z\cos{\ch}
  + \tilde{b}_X\sin{\ch}\cos{\Om t} 
  + \tilde{b}_Y\sin{\ch}\sin{\Om t}            
 \quad ,
  \nonumber \\
\tildec_q&=&
 \tildec_Q (\frac{3}{2}\cos{^2\ch}-\half)
  \nonumber \\
 && - \frac 32\tildec_{Q,X}\sin{2\ch}\cos{\Om t}
  -\frac 32\tildec_{Q,Y}\sin{2\ch}\sin{\Om t}
  \nonumber \\
 && - \frac 32 \tildec_-\sin{^2\ch}\cos{2\Om t}
  - \frac 32 \tildec_{XY}\sin{^2\ch}\sin{2\Om t}
 \quad ,
  \nonumber \\
\tilde{d}_3 &=& \tilde{d}_Z\cos{\ch} 
  + \tilde{d}_X\sin{\ch}\cos{\Om t} 
  + \tilde{d}_Y\sin{\ch}\sin{\Om t}  
 \quad ,
  \nonumber \\
\tildeg_d &=&
 \tildeg_{D,Z} \cos{\ch}
 + \tildeg_{D,X} \sin{\ch}\cos{\Om t}
 + \tildeg_{D,Y} \sin{\ch}\sin{\Om t}
 \quad ,
  \nonumber \\
\tildeg_q &=& 
 \tildeg_Q (\frac{3}{2}\cos{^2\ch}-\half)
  \nonumber \\
 && - \frac 32\tildeg_{Q,X}\sin{2\ch}\cos{\Om t}
  - \frac 32\tildeg_{Q,Y}\sin{2\ch}\sin{\Om t}
  \nonumber \\
 && - \frac 32 
  \tildeg_- \sin{^2\ch}\cos{2\Om t}
  - \frac 32 
  \tildeg_{XY} \sin{^2\ch}\sin{2\Om t}
 \quad .
\label{transformbdcg}
\eea
Note that 
$\tildeb_3$, $\tilded_3$, and $\tildeg_d$
involve constant pieces and ones varying with the sidereal frequency $\Om$,
while the others also have terms varying with the
semi-sidereal frequency $2\Om$.
Note also that the parameters
$\tildeb_Z$,
$\tildec_Q$,
$\tilded_Z$,
$\tildeg_{D,Z}$,
$\tildeg_Q$
appear only in time-independent terms,
and they therefore are unconstrained by clock-comparison experiments.

Substituting the above into the expression \rf{shift}
for the energy-level shift gives
\bea
\bra{F,m_F} \pr{h} \ket{F,m_F} &=& E_0
+E_{1X}\cos{\Om t}+E_{1Y}\sin{\Om t}
   \nonumber \\
 &&+ E_{2X}\cos{2\Om t}+E_{2Y}\sin{2\Om t}
\quad .
\nonumber \\
\label{tdatoms}
\eea
The energy $E_0$ is constant in time 
and is therefore irrelevant for clock-comparison experiments.
The four other energies are defined by
\bea
E_{1X} &:=& 
   \hmf \sin{\ch}
\sum_{w} 
   ( \be_w \tilde{b}_X^w 
         +\de_w \tilde{d}_X^w 
         +\ka_w \tildeg_{D,X}^w ) 
   \nonumber \\
   && - 
   \frac 32 \tmf \sin{2\ch}
\sum_{w} 
   ( \ga_w \tildec_{Q,X}^w  
         +\la_w \tildeg_{Q,X}^w )
\quad , 
 \nonumber \\
E_{1Y} &:=& 
   \hmf \sin{\ch}
\sum_{w} 
   ( \be_w \tilde{b}_Y^w
        +\de_w \tilde{d}_Y^w 
        +\ka_w \tildeg_{D,Y}^w ) 
 \nonumber \\
   && -
   \frac 32 \tmf \sin{2\ch}
\sum_{w} 
   ( \ga_w \tildec_{Q,Y}^w 
         +\la_w \tildeg_{Q,Y}^w )
\quad , 
 \nonumber \\
E_{2X} &:=& 
   -\frac{3}{2} \tmf \sin{^2\ch}
\sum_{w}
   ( \ga_w \tildec_-^w + \la_w \tildeg_-^w ) 
\quad , 
 \nonumber \\
E_{2Y} &:=& 
   -\frac{3}{2} \tmf \sin{^2\ch}
\sum_{w}
   ( \ga_w \tildec_{XY}^w + \la_w \tildeg_{XY}^w )
\quad .
\label{define_omegas}
\eea
For clock-comparison experiments,
the signal is typically a time variation in a frequency.
In the context of the present theoretical framework,
this is determined by the difference between 
two energy-level shifts of the form Eq.\ \rf{tdatoms}.

The reader should note that the component of the shift 
in Eq.\ \rf{tdatoms}
varying with the sidereal frequency $\Om$ 
is determined by operators producing
both dipole and quadrupole energy shifts.
The issue of the multipolarity of the energy shift,
which is governed by the rotation properties of the 
Lorentz-violating perturbations in the laboratory frame,
is somewhat different from the issue of the periodicity of the signals in 
clock-comparison experiments,
which is governed also by factors associated with the Earth's rotation.
The relatively simple correspondences
sometimes found in the literature 
between the multipolarity of the energy shift 
and the periodicity of the signal or the effects on the spectrum 
are invalid in the general case
\cite{fn3}.

The use of the nonrelativistic transformation
\rf{coordmatrix}
between the nonrotating and laboratory frames means that 
contributions from nonrotating-frame time components of the
parameters for Lorentz violation are absent.
In a more exact treatment,
these quantities would be present
but suppressed by a factor of order $ 10^{-6}$.
For example,
$\tildeb_3$ strictly also depends slightly
on the nonrotating-frame timelike component $b_0$.
This means that some bounds on certain time components of the
parameters could in principle be obtained.
However, these bounds would be much weaker than 
the ones considered here.
Moreover,
an accurate treatment would also require inclusion
of some of the other subleading effects mentioned 
at the beginning of this section.

\section{Application}
\label{application}

This section applies
the theoretical framework of the previous section
to existing and future clock-comparison experiments.
The limits attained in the original experiments 
of Hughes {\it et al.} \cite{hughes} and Drever \cite{drever}
have been improved by many orders of magnitude
in recent years.
In the first part of this section,
we focus our attention on results 
from the clock-comparison experiments performed by
Prestage {\it et al.} \cite{prestage},
Lamoreaux {\it et al.} \cite{lamoreaux},
Chupp {\it et al.} \cite{chupp},
and Berglund {\it et al.} \cite{berglund}.
The theory presented in section \ref{theory}
can be used to extract 
from each of these experiments
one or more bounds on combinations 
of parameters for Lorentz violation.
In the second part of this section,
we present some considerations relevant
to possible future experiments.

For purposes of discussion,
it is useful to format all the bounds in a unified way.
In effect,
each experiment observes the frequency 
of one atomic or ionic species $A$
relative to a reference frequency in another species $B$,
producing one or more bounds
on possible sidereal or fractional-sidereal variations
as the Earth rotates.
Within the present framework,
the effect of Lorentz violations on these frequencies
can be deduced from the perturbative energy shifts
given in Eq.\ \rf{tdatoms}.
Some comments about this procedure are provided in Appendix A.

We find that each bound from each experiment
fits one of the following forms:
\widetext
\top{-2.8cm}
\hglue -1 cm
\bea
&& 
 \Big| 
 \sum_{w} 
  [ u_0^A( \be_w^A \tildeb_X^w + \de_w^A \tilded_X^w 
        + \ka_w^A \tildeg_{D,X}^w )
   + u_1^A( \ga_w^A \tildec_{Q,X}^w 
           + \la_w^A \tildeg_{Q,X}^w ) ]
 \nonumber\\
  &&
  \qquad\qquad\qquad\qquad
  - v\sum_{w} 
    [ u_0^B( \be_w^B \tildeb_X^w + \de_w^B \tilded_X^w 
           + \ka_w^B \tildeg_{D,X}^w )
     + u_1^B( \ga_w^B \tildec_{Q,X}^w 
           + \la_w^B \tildeg_{Q,X}^w ) ]
  \Big| 
\lsim 2\pi\ve_{1,X} 
\quad , 
   \nonumber \\
&& 
 \Big| 
 \sum_{w} 
  [ u_0^A( \be_w^A \tildeb_Y^w + \de_w^A \tilded_Y^w 
        + \ka_w^A \tildeg_{D,Y}^w )
   + u_1^A( \ga_w^A \tildec_{Q,Y}^w 
           + \la_w^A \tildeg_{Q,Y}^w ) ]
 \nonumber\\
  &&
  \qquad\qquad\qquad\qquad
  - v\sum_{w} 
    [ u_0^B( \be_w^B \tildeb_Y^w + \de_w^B \tilded_Y^w 
           + \ka_w^B \tildeg_{D,Y}^w )
     + u_1^B( \ga_w^B \tildec_{Q,Y}^w 
           + \la_w^B \tildeg_{Q,Y}^w ) ]
  \Big| 
\lsim 2\pi\ve_{1,Y}
\quad , 
   \nonumber \\
&& 
 \Big| \sum_{w} 
  u_2^A
  ( \ga_w^A \tildec_-^w 
   +\la_w^A \tildeg_-^w )
  -v \sum_{w}
  u_2^B
  ( \ga_w^B \tildec_-^w 
   +\la_w^B \tildeg_-^w )
 \Big|
\lsim 2\pi \ve_{2,-}
\quad ,
   \nonumber \\
&&
 \Big| \sum_{w} 
  u_2^A
  ( \ga_w^A \tildec_{XY}^w
   +\la_w^A \tildeg_{XY}^w )
  -v \sum_{w}
  u_2^B
  ( \ga_w^B \tildec_{XY}^w 
   +\la_w^B \tildeg_{XY}^w )
 \Big| 
\lsim 2\pi \ve_{2,XY}
\quad .
\label{generalbound}
\eea
\bottom{-2.7cm}
\narrowtext
\noindent
Here,
the coefficients $u_0$, $u_1$, $u_2$, and $v$
contain the dependences on quantities such as 
$\hmf$, $\tmf$, $\ch$, and gyromagnetic ratios.
For example,
if an atom or ion $W$ undergoes a transition 
$\ket{F, \pr{m}_F}\to\ket{F,m_F}$,
then  
$u_0 = (\pr{\hmf}-\hmf) \sin{\ch}$,
$u_1 = -\frac{3}{2} (\pr{\tmf}-\tmf) \sin{2\ch}$,
and $u_2 = -\frac{3}{2} (\pr{\tmf}-\tmf) \sin{^2\ch}$.
The parameter $v=g_A/g_B$ is the ratio 
of gyromagnetic ratios for the species $A$ and $B$.
Also,
the experimental bounds on the amplitudes 
of frequency shifts are denoted by 
$\ve_{1,X}$,
$\ve_{1,Y}$,
$\ve_{2,-}$, 
$\ve_{2,XY}$, 
corresponding to sidereal or semi-sidereal variations as 
$\cos \Om t$,
$\sin \Om t$, 
$\cos 2\Om t$,
$\sin 2\Om t$, 
respectively.
The other quantities are defined earlier in the text.
For example,
the factors $\be_w$, $\ga_w$, $\de_w$, $\ka_w$, and $\la_w$ 
are those given in Eq.\ \rf{bgdkl},
with subscripts according to the particle species
and superscripts according to the atomic or ionic species.
The components of the parameters for Lorentz violation
are those for the nonrotating frame
$(\X,\Y,\Z)$.
Note that the second of the equations is the same as the first
but with the replacement $X\to Y$,
reflecting the arbitrariness of the choice 
of $X$ and $Y$ axes in the $XY$ plane.
The form of the third and fourth of the above equations 
also reflects this arbitrariness.

The values of all relevant coefficients for each 
of the experiments we consider are summarized in Table 1.
The first few rows of this table identify the experiment and provide
information about the atoms or ions used. 
We denote the nuclear spin by $I$, 
the proton number by $Z$,
and the neutron number by $N$.
The nucleon determining the ground-state properties of the nucleus
according to the nuclear Schmidt model \cite{schmidt,blatt}
is specified,
along with its assignment of orbital and total angular momenta.
Fifteen rows are devoted to the values of the coefficients
$\be_w$, $\ga_w$, $\de_w$, $\ka_w$, and $\la_w$
for each atom or ion.
In these rows,
a dash indicates that the (possibly nonzero) coefficient 
is irrelevant for the experiment.
Values in brackets are results obtained 
within the Schmidt model,
with $K_w$ defined by $K_w := \vev{p^2}_w/m_w^2$
as in subsection \ref{hamiltonian}.
These values are to be trusted only where 
the model is known to give reliable results.
Any zero values in brackets are probably unreliable 
because they are likely to be nonzero 
in more realistic nuclear models.
Zero values without brackets are a consequence 
of the Wigner-Eckart theorem applied to closed shells,
according to the discussion at the end of subsection \ref{coexva},
and therefore depend on fewer assumptions.
The following four rows specify the coefficients 
$u_0$, $u_1$, $u_2$, $v$.
A dash indicates the coefficient is irrelevant for the experiment.
Two rows provide approximate values for 
the experimental sidereal and semi-sidereal bounds obtained.
In the final row, 
a list is provided of the parameters for Lorentz violation
constrained by the experiment according to this analysis.

The table reveals a crucial point:
the published experiments are all inequivalent,
in the sense that they bound different 
linear combinations of parameters for Lorentz violation.
In some cases, 
such as the experiments of 
Chupp {\it et al.} \cite{chupp}
and Berglund {\it et al.} \cite{berglund},
there is no overlap at all
among the set of parameters for Lorentz violation
appearing in the bounds.
In other cases,
such as the experiments of 
Prestage {\it et al.} \cite{prestage}
and Lamoreaux {\it et al.} \cite{lamoreaux},
there is substantial overlap among
the parameters for Lorentz violation involved
but they appear in different linear combinations. 

In the context of the Schmidt model,
the Prestage {\it et al.},
Lamoreaux {\it et al.},
and Chupp {\it et al.}
experiments are sensitive only to 
(different sets of)
parameters for Lorentz violation involving the neutron,
$w \equiv n$.  
In contrast,
the experiment of Berglund {\it et al.} 
involves all three particle species 
because the $\hg$ atom is sensitive to parameters 
for Lorentz violation with $w \equiv n$
and the $\cs$ atom is sensitive to ones with $w \equiv p, e$.
None of the experiments considered
place any bound on the quantities
$\tildec_{Q,J}^e$,
$\tildec_-^e$,
$\tildec_{XY}^e$,
$\tildeg_{Q,J}^e$,
$\tildeg_-^e$,
$\tildeg_{XY}^e$,
while within the Schmidt model no bounds are placed on
$\tildec_{Q,J}^p$,
$\tildec_{XY}^p$,
$\tildec_-^p$,
$\tildeg_{Q,J}^p$,
$\tildeg_{XY}^p$,
$\tildeg_-^p$,
$\tildeg_{Q,J}^n$,
$\tildeg_{XY}^n$,
$\tildeg_-^n$.
Note that some of these quantities can be bounded 
in other kinds of experiments \cite{bkr,gg}.

A more accurate nuclear model 
would be likely to introduce dependence on parameters 
with $w \equiv p$ for all atoms and ions in these experiments 
except the H maser used by Prestage {\it et al.}
and the $\he$ reference used by Chupp {\it et al.}
Thus,
the true bounds from all four experiments are likely to 
involve parameters for more than one species $w$.
For the experiment of Berglund {\it et al.},
the true dependence on parameters with $w \equiv p$ 
might even come primarily from the $\hg$ atom
because the contribution from the $\cs$ atom is
suppressed by its relatively large gyromagnetic ratio,
reflected in Table 1 by the small size of the coefficient $v$.

The numerical values of the bounds obtained 
in all these experiments are impressive
and represent sensitivity to Planck-scale physics.
In contrast,
the relatively complicated form 
of the linear combinations bounded
and the theoretical issues involved 
in accurately determining the various coefficients
make it difficult to establish definitively
which portion of the parameter space is
in fact being excluded.
It is therefore of some interest to speculate 
about the implications of these experiments
under different scenarios
that lead to theoretically cleaner bounds.

\widetext
\medskip
\begin{center}
{Table 1: Coefficients for the bounds \rf{generalbound}
for various experiments.}
\end{center}
\begin{center}

\begin{tabular}{||c||c|c||c|c||c|c||c|c||}
\hline
\hline
  & \multicolumn{2}{c||}{ \quad Prestage {\it et al.} [3] \quad }
  & \multicolumn{2}{c||}{ \quad Lamoreaux {\it et al.} [4] \quad }
  & \multicolumn{2}{c||}{ \quad Chupp {\it et al.} [5] \quad }
  & \multicolumn{2}{c||}{ \quad Berglund {\it et al.} [6] \quad } \\
\hline
\hline
   & $A$  & $B$  & $A$  & $B$  & $A$  & $B$  & $A$  & $B$ \\
\hline 
   & $\ber$  & ${^1}$H  & $\hgg$  & $\hg$   & $\ne$  & $\he$  
  & $\hg$  & $\cs$ \\ 
\hline
 $I$   & $3/2$  & $1/2$  & $3/2$  & $1/2$   & $3/2$  & $1/2$ & $1/2$ & $7/2$
\\ 
 $Z$  & $4$  & $1$  & $80$  & $80$   & $10$  & $2$ & $80$ & $55$ \\ 
 $N$  & $5$  & $0$  & $121$  & $119$   & $11$  & $1$ & $119$ & $78$ \\ 
\hline
 Schmidt & $p_{3/2}$  & $s_{1/2}$  & $p_{3/2}$  
   & $p_{1/2}$   & $p_{3/2}$  & $s_{1/2}$ & $p_{1/2}$ & $g_{7/2}$ \\ 
 nucleon  & $n$  & $p$  & $n$  & $n$   & $n$  & $n$ & $n$ & $p$ \\ 
\hline
\hline
 $\be_p$  & $[0]$  & $-$  & $[0]$  & $[0]$     & $-$  & $-$  
  & $[0]$  
  & $[\frac{7}{9}]$ \\
 $\ga_p$ & $[0]$  & $-$  & $[0]$  & $0$     & $[0]$  & $0$  
  & $-$  & $-$ \\
 $\de_p$ & $[0]$  & $-$  & $[0]$  & $[0]$     & $-$  & $-$  
  & $[0]$  & $[-\frac{7}{33}\kinp]$ \\
 $\ka_p$ & $[0]$  & $-$  & $[0]$  & $[0]$     & $-$  & $-$  
  & $[0]$  & $[\frac{28}{99}\kinp]$ \\
 $\la_p$ & $[0]$  & $-$  & $[0]$  & $0$     & $[0]$  & $0$  
  & $-$  & $-$ \\ 
\hline
 $\be_n$  & $[-1]$  & $-$  & $[-1]$  & $[1/3]$  & $-$    & $-$  
  & $[1/3]$  & $[0]$ \\
 $\ga_n$ & $[-\frac{1}{15}\kinn]$  & $-$  & $[-\frac{1}{15}\kinn]$ 
  & $0$     & $[-\frac{1}{15}\kinn]$  & $0$  & $-$  & $-$ \\
 $\de_n$ & $[\frac{1}{5}\kinn]$  & $-$  & $[\frac{1}{5}\kinn]$   
  & $[-\frac{1}{5}\kinn]$    & $-$ & $-$  & $[-\frac{1}{5}\kinn]$  
  & $[0]$ \\
 $\ka_n$ & $[-\frac{2}{5}\kinn]$  & $-$  & $[-\frac{2}{5}\kinn]$  
  & $[\frac{1}{15}\kinn]$   & $-$  & $-$  
  & $[\frac{1}{15}\kinn]$  & $[0]$ \\
 $\la_n$ & $[0]$  & $-$  & $[0]$  & $0$  & $[0]$     & $0$  
  & $-$  & $-$ \\
\hline
 $\be_e$   & $-$  & $-$  & $0$  & $0$  & $-$  & $-$  & $0$     
  & $[-1]$ \\
 $\ga_e$ & $-$  & $-$  & $0$  & $0$  & $0$  & $0$     & $-$  
  & $-$ \\
 $\de_e$ & $-$  & $-$  & $0$  & $0$  & $-$  & $-$     & $0$  
  & $[\frac{1}{3}\kine]$ \\
 $\ka_e$ & $-$  & $-$  & $0$  & $0$  & $-$  & $-$     & $0$  
  & $[-\frac{1}{3}\kine]$ \\
 $\la_e$ & $-$  & $-$  & $0$  & $0$  & $0$  & $0$     & $-$  
  & $-$ \\
\hline
\hline
 $u_0$ 
  & $-0.61$ & --   
  & $\frac{2}{3}$ & $2$   
  & -- & --
  & $2$ & $\frac{1}{4}$ \\
 $u_1$
  & $2.16$ & -- 
  & 0 & 0
  & -- & -- 
  & 0 & 0 \\
 $u_2$
  & $-2.54$ & -- 
  & $-3$ & $0$ 
  & $-3$ & -- 
  & -- & -- \\
\hline
 $v$ 
  & \multicolumn{2}{c||}{ -- } 
  & \multicolumn{2}{c||}{ $-0.37$ } 
  & \multicolumn{2}{c||}{ -- } 
  & \multicolumn{2}{c||}{ $2.2\times 10^{-3}$ } \\
\hline
 $\ve_{1,X}$, $\ve_{1,Y}$ 
  & \multicolumn{2}{c||}{ $\simeq 100\ \mu$Hz }
  & \multicolumn{2}{c||}{ $\simeq 1\ \mu$Hz } 
  & \multicolumn{2}{c||}{ no bound }
  & \multicolumn{2}{c||}{ $\simeq 100\ $nHz } \\
 $\ve_{2,-}$, $\ve_{2,XY}$ 
  & \multicolumn{2}{c||}{ $\simeq 100\ \mu$Hz }
  & \multicolumn{2}{c||}{ $\simeq 1\ \mu$Hz }
  & \multicolumn{2}{c||}{ $\simeq 1\ \mu$Hz }
  & \multicolumn{2}{c||}{ no bound } \\
\hline
  & \multicolumn{2}{c||}{ }
  & \multicolumn{2}{c||}{ }
  & \multicolumn{2}{c||}{ }
  & \multicolumn{2}{c||}{ } \\ 
 Sensitivity
  & \multicolumn{2}{c||}{ $[\tilde{b}_J^n,\tilde{d}_J^n,
       \tildeg_{D,J}^n,$ }
  & \multicolumn{2}{c||}{ $[\tilde{b}_J^n,\tilde{d}_J^n,
       \tildeg_{D,J}^n,\tildec_-^n,\tildec_{XY}^n]$ }
  & \multicolumn{2}{c||}{ $[\tildec_-^n,\tildec_{XY}^n]$ }
  & \multicolumn{2}{c||}{ $[\tilde{b}_J^p,\tilde{d}_J^p,
       \tildeg_{D,J}^p,\tilde{b}_J^n,\tilde{d}_J^n,\tildeg_{D,J}^n,$ } \\ 
 $(J=X,Y)$
  & \multicolumn{2}{c||}{ $\tildec_{Q,J}^n,\tildec_-^n,\tildec_{XY}^n]$ }
  & \multicolumn{2}{c||}{  }
  & \multicolumn{2}{c||}{ }
  & \multicolumn{2}{c||}{ $\tilde{b}_J^e,\tilde{d}_J^e,\tildeg_{D,J}^e]$ }
\\
  & \multicolumn{2}{c||}{ }
  & \multicolumn{2}{c||}{ }
  & \multicolumn{2}{c||}{ }
  & \multicolumn{2}{c||}{ } \\
\hline \hline
\end{tabular}

\end{center}
\narrowtext

One plausible assumption is that effects from the first
atomic or ionic species $A$ would be unlikely 
to cancel completely the effects from the second species $B$
in Eq.\ \rf{generalbound}.
This assumption would permit (perhaps numerically weaker) bounds
to be placed on somewhat simpler combinations of parameters.
A stronger extension of this assumption might also be adopted
to the effect that for a single species 
exact cancellations are unlikely among different terms in the sums
appearing in Eq.\ \rf{generalbound}.
If this stronger assumption also holds,
then for a given experimental bound 
the numerical value could be applied
to each term in the sum,
yielding plausible (but not definite)
constraints on each of the parameters for Lorentz violation
appearing in Eq.\ \rf{generalbound}.

To gain some insight into the implications of these assumptions,
one can examine the bounds that would follow 
within the additional approximation of the Schmidt model.
Then,
the sole experiment bounding electron or proton parameters
for Lorentz violation is that of 
Berglund {\it et al.},
which constrains only 
$\tildeb_J^w$,
$\tilded_J^w$,
and $\tildeg_{D,J}^w$
for $J=X,Y$.
Also within these assumptions,
the earlier experiments of
Prestage {\it et al.},
Lamoreaux {\it et al.},
and Chupp {\it et al.}
are insensitive to electron or proton parameters
for Lorentz violation,
but instead they have sensitivity to neutron parameters 
beyond the ones constrained by Berglund {\it et al.}
For example,
the experiment of Prestage {\it et al.}
is the only one that constrains $\tildec_{Q,J}^n$.

Within these strong (and questionable) assumptions,
approximate numerical bounds can be obtained by
using dimensional estimates for the quantities $K_w$.
As in subsection \ref{hamiltonian},
we take the crude values $K_p \approx K_n \simeq 10^{-2}$
and $K_e \simeq 10^{-5}$.
Table 2 presents the results of this estimate.
In a given row,
numerical values in brackets are estimated order-of-magnitude bounds 
in GeV obtained within the Schmidt model
assuming that all the parameters for Lorentz violation are zero except
for the one specific to that row.
The symbol $\star$ indicates that no experimental bound is placed 
according to the Schmidt model
but that a bound probably would emerge from a 
more realistic nuclear model.
A dash indicates insensitivity to the specified parameter 
for Lorentz violation.
The values obtained
represent bounds on different parameters for Lorentz violation
varying over about eight orders of magnitude,
with the sharpest being a constraint
on $\tildeb_J^n$ of about $10^{-30}$ GeV.
Although the approximations made imply that the specific numerical bounds
listed in Table 2 are unreliable, 
in certain cases perhaps within several orders of magnitude, 
they nonetheless provide a gauge of the present impressive sensitivity 
of these experiments to the relevant parameters for Lorentz violation.

Still another speculation one might entertain,
in addition to the above assumptions,
is to suppose that cancellations 
are unlikely within each of the linear combinations
in Eq.\ \rf{bdcgtildenonrot}.
If this were valid,
then one could apply the numerical experimental bounds
to deduce constraints on the original parameters 
for Lorentz violation appearing in the 
QED and standard-model extensions
instead of merely constraining some of the combinations
given in Eq.\ \rf{bdcgtildenonrot}.
However, 
this assumption is difficult to justify because 
it is likely that the various parameters in the QED extension
are related through the underlying theory,
perhaps in a relatively simple way,
so significant cancellations may well occur.
For example,
a simple relation among certain parameters 
for CPT and Lorentz violation 
is known to occur in an anomaly-free underlying theory,
and in this case it indeed eliminates 
the sensitivity of some experiments 
(on cosmological birefringence) 
to certain parameters for Lorentz violation 
\cite{cksm,jk}.

\medskip
\begin{center}
{Table 2: Crude order-of-magnitude bounds in GeV 
on parameters for Lorentz violation.}
\end{center}
\begin{center}

\begin{tabular}{|c|c|c|c|c|}
 \hline
Expt. 	& [3] 	& [4] 
 	& [5] 	& [6]  
 \\ \hline
$\tildeb_J^p$		& $\star$ 	& $\star$ 	& -- 	&
$[10^{-27}]$	\\
$\tildec_{Q,J}^p$	& $\star$ 	& --       	& -- 	& -- 	\\
$\tildec_-^p$		& $\star$ 	& $\star$	& $\star$	&
-- 	\\
$\tildec_{XY}^p$	& $\star$ 	& $\star$	& $\star$	&
-- 	\\
$\tilded_J^p$		& $\star$ 	& $\star$ 	& -- 	&
$[10^{-25}]$	\\ 
$\tildeg_{D,J}^p$	& $\star$ 	& $\star$ 	& -- 	&
$[10^{-25}]$	\\
$\tildeg_{Q,J}^p$	& $\star$ 	& -- 	& -- 	&  -- 	\\
$\tildeg_-^p$		& $\star$	& $\star$	& $\star$	&
-- 	\\
$\tildeg_{XY}^p$ 	& $\star$	& $\star$	& $\star$	&
-- 	\\
 \hline
$\tildeb_J^n$		& $[10^{-27}]$	& $[10^{-29}]$	& -- 	&
$[10^{-30}]$	\\
$\tildec_{Q,J}^n$	& $[10^{-25}]$	& -- 	& -- 	& -- 	\\
$\tildec_-^n$		& $[10^{-25}]$	& $[10^{-27}]$	& $[10^{-27}]$	&
-- 	\\
$\tildec_{XY}^n$ 	& $[10^{-25}]$	& $[10^{-27}]$	& $[10^{-27}]$	&
-- 	\\
$\tilded_J^n$		& $[10^{-25}]$	& $[10^{-26}]$	& -- 	&
$[10^{-28}]$	\\ 
$\tildeg_{D,J}^n$	& $[10^{-25}]$	& $[10^{-27}]$	& -- 	&
$[10^{-28}]$	\\
$\tildeg_{Q,J}^n$	& $\star$ 	& -- 	& -- 	& -- 	\\
$\tildeg_-^n$		& $\star$	& $\star$	& $\star$	&
-- 	\\
$\tildeg_{XY}^n$ 	& $\star$	& $\star$	& $\star$	&
-- 	\\
 \hline
$\tildeb_J^e$		& -- 	& -- 	& -- 	& $[10^{-27}]$	\\
$\tildec_{Q,J}^e$	& -- 	& -- 	& -- 	& -- 	\\
$\tildec_-^e$		& -- 	& -- 	&  --	& -- 	\\
$\tildec_{XY}^e$ 	& -- 	& -- 	&  --	& -- 	\\
$\tilded_J^e$		& -- 	& -- 	& -- 	& $[10^{-22}]$	\\ 
$\tildeg_{D,J}^e$	& -- 	& -- 	& -- 	& $[10^{-22}]$	\\
$\tildeg_{Q,J}^e$	& -- 	& -- 	& -- 	& -- 	\\
$\tildeg_-^e$		& -- 	&  --	& -- 	& -- 	\\
$\tildeg_{XY}^e$ 	& -- 	&  --	& -- 	& -- 	\\
 \hline
\end{tabular}

\end{center}

Despite the relatively complicated form of the bounds \rf{generalbound}
and the theoretical issues in calculating the associated coefficients,
the exceptional precision attained makes clock-comparison experiments 
of great interest.
In principle,
from a theoretical perspective
a particularly attractive type of bound would be one 
that is both calculable and clean. 
Here,
\it calculable \rm
refers to the reliability with which the various coefficients
can be theoretically established,
and
\it clean \rm 
refers to the number of different parameters
for Lorentz violation involved in the bound
(the fewer the cleaner).
It is interesting to examine the extent to which 
calculable and clean bounds from clock-comparison experiments
are theoretically possible. 

For a bound to be calculable with the methods adopted here,
reliable wave functions would need to be obtained 
for the atoms or ions used in the experiment.
The complexity of nuclear matter 
typically makes this a challenging task.
One possibility is to consider atoms or ions 
involving very few particles,
so that a detailed calculation has an improved chance of accuracy.

Among the simplest substances is hydrogen.
The well-developed theoretical understanding 
of the hydrogen wave function
makes it a prime candidate for a substance
that would produce a calculable bound.
Various experiments with hydrogen atoms and ions
(H$^-$ and the proton)
and with antihydrogen
have been investigated 
in the context of the present theoretical framework
in Refs.\ \cite{bkr,bkr2}.
A hydrogen maser was used as a reference 
in the clock-comparison experiment of Prestage {\it et al.}
The conventional H-maser line 
involves atomic states with $m_F=0$
and hence is insensitive
to the parameters for Lorentz violation,
which simplifies the resulting experimental bounds.
Other ground-state hyperfine lines in hydrogen 
involve states with $m_F=\pm 1$,
and they depend on parameters for Lorentz violation
according to Eq.\ (5) of Ref.\ \cite{bkr2}.
At leading order,
the sidereal variations of these hyperfine lines are governed by
the strength and orientation of the applied magnetic field
and the combinations $\tildeb^e_J \pm \tildeb^p_J$.
Experiments searching for this dependence,
perhaps with trapped hydrogen or a hydrogen maser,
have the potential to yield calculable bounds. 
Moreover,
since no neutrons are involved,
only electron and proton parameters for Lorentz violation would appear.

Among other atoms and ions involving relatively 
few particles is deuterium.
To our knowledge,
no clock-comparison experiments have been performed with deuterium.
However,
an experiment by Wineland and Ramsey \cite{wineland}
studied transition frequencies in a deuterium maser.
The transition $\ket{F,m_F}\equiv \ket{\frac 3 2,\frac 1 2} 
\to \ket{\frac 1 2,-\frac 1 2}$
was compared when the (weak) applied magnetic field 
was parallel and antiparallel to the Earth's gravitational field.
The result constrains the possible frequency difference
to about 200 $\mu$Hz. 
This experiment was performed to bound 
the gravitational dipole moment of the deuteron,
for which the orientation of the magnetic field 
relative to a nonrotating frame is irrelevant,
and a useful bound on Lorentz violation is difficult to extract from it.
However,
the possibility of using a deuterium maser 
as one or both clocks
in an experiment to bound Lorentz violation is worth consideration
because the neutral deuterium atom is relatively well understood
from a theoretical viewpoint.
It consists of a deuteron ($I=1$) and an electron ($l_j=s_{1/2}$),
which can combine into states with $F=3/2$ or $F=1/2$
\cite{nfr}.
Explicit forms of the deuteron wave function exist 
\cite{deuteron}
and could in principle be used to calculate the coefficients
in Eq.\ \rf{bgdkl} for each particle species.
The deuterium atom therefore provides another example of a substance 
that could produce a calculable bound.
However,
the deuterium energy-level shifts
are sensitive to parameters for Lorentz violation 
involving all particle species,
so any bound attained is unlikely to be clean
in the sense defined above.

Perhaps the ideally clean clock-comparison experiment would be one 
for which one atomic or ionic species 
is insensitive to Lorentz violation
and the other is sensitive to only one 
of the minimal set of parameters for Lorentz violation 
$\tildeb_J^w$, 
$\tildec_{Q,J}^w$, 
$\tildec_-^w$, 
$\tildec_{XY}^w$, 
$\tilded_J^w$,
$\tildeg_{D,J}^w$,
$\tildeg_{Q,J}^w$,
$\tildeg_-^w$,
$\tildeg_{XY}^w$ 
discussed in the analysis of section \ref{theory}.
In practice, 
however, 
this ideal is unlikely to be attainable.
Insensitive systems such as the hydrogen maser do exist,
and in principle an insensitive system 
could be obtained for any substance 
by aligning the applied magnetic field
with the Earth's rotation axis.
However,
sensitivity to only one of the minimal parameters
for Lorentz violation is difficult to achieve.
For example,
if a nonzero effect on the energy levels of an atom or ion
involves $\tildeb_J^w$ 
then it also involves $\tilded_J^w$.
Comparisons of bounds from different experiments
may permit the extraction of a bound on a single 
parameter for Lorentz violation,
but the issue of the calculability of the coefficients
would again play an important role.

An interesting option for improving both the calculability 
and the cleanliness of bounds is to consider atoms or 
(positive or negative) ions 
for which there is reason to believe that the energy shifts depend
solely or largely on a single valence particle $w$.
The presence of only one relevant particle can simplify calculations,
and substances of this type would also be relatively clean
because only those parameters for Lorentz violation for a 
particular species $w$ would be involved in the bound.

It is relatively straightforward to identify atoms or ions for which 
the special species $w$ is an electron,
since it suffices to use substances of nuclear spin zero 
(or nuclear spin $\half$, 
if only bounds on quadrupole energy shifts 
for the electron are considered).
These bounds would be of definite interest,
even if the precision attainable were less than 
in experiments with hyperfine transitions.

For the case where the special species $w$ is a nucleon,
one can generate a list of nuclei
for which one might theoretically expect 
relatively calculable and clean bounds.
We consider here substances
for which dipole and quadrupole energy shifts
depend only on one nucleon species
and where there is reason to believe that 
only one valence nucleon determines the nuclear sensitivity
to Lorentz violation.
Table 3 provides a list of naturally abundant isotopes
satisfying these criteria.
The table has been prepared using only a relatively small set 
of assumptions about nuclear properties:
spin-$\half$ nuclei are assumed 
to be insensitive to quadrupole effects,
while those with a closed shell for a given nucleon species
are assumed to be insensitive to parameters for Lorentz violation
for that species.
Note in particular that the Schmidt model has \it not \rm been used.
To avoid complicating the bounds through 
sensitivity to electron parameters for Lorentz violation,
experiments on any of the substances listed would need to involve 
suitable electronic configurations without 
Lorentz-violating contributions to the 
relevant transition frequencies.
The reader is warned that the table makes no allowance for
possible experimental difficulties involved in using these substances.

\widetext
\medskip
\begin{center}
{Table 3: Substances with sensitivity to 
parameters for Lorentz violation for a single particle species.}
\end{center}
\begin{center}

\begin{tabular}{|c|c|c|c|c|cc|cc|c|||c|c|c|c|c|cc|cc|c|} 
\hline
\multicolumn{10}{|c|||}{Proton Sensitivity Only} &
   \multicolumn{10}{c|}{Neutron Sensitivity Only} 
\\ \hline
 & $A$ & $Z$ & $N$ & $I$ & $D_p$ & $Q_p$ & $D_n$ & $Q_n$ & &
 & $A$ & $Z$ & $N$ & $I$ & $D_p$ & $Q_p$ & $D_n$ & $Q_n$ & 
 \\ \hline 
H & 1 & 1 & 0 & 1/2 & Y & -- & -- & -- & $\star\star$ &
n & 1 & 0 & 1 & 1/2 & -- & -- & Y & -- & $\star\star$
 \\ \hline 
N & 15 & 7 & 8 & 1/2 & Y & -- & -- & -- & $\star$ &
He & 3 & 2 & 1 & 1/2 & -- & -- & Y & -- & $\star\star$
 \\ \hline 
P & 31 & 15 & 16 & 1/2 & Y & -- & -- & -- & &
C & 13 & 6 & 7 & 1/2 & -- & -- & Y & -- &
 \\ \hline 
Y & 89 & 39 & 50 & 1/2 & Y & -- & -- & -- & $\star$ &
Si & 29 & 14 & 15 & 1/2 & -- & -- & Y & -- &
 \\ \hline 
Rh & 103 & 45 & 58 & 1/2 & Y & -- & -- & -- & &
Sn & 115 & 50 & 65 & 1/2 & -- & -- & Y & -- & $\star$
 \\ \hline 
Tm & 169 & 69 & 100 & 1/2 & Y & -- & -- & -- & &
Sn & 117 & 50 & 67 & 1/2 & -- & -- & Y & -- & $\star$
 \\ \hline 
B & 11 & 5 & 6 & 3/2 & Y & Y & -- & -- & &
Sn & 119 & 50 & 69 & 1/2 & -- & -- & Y & -- & $\star$
 \\ \hline 
Al & 27 & 13 & 14 & 5/2 & Y & Y & -- & -- & &
Yb & 171 & 70 & 101 & 1/2 & -- & -- & Y & -- &
 \\ \hline 
Cl & 37 & 17 & 20 & 3/2 & Y & Y & -- & -- & $\star$ &
Pb & 207 & 82 & 125 & 1/2 & -- & -- & Y & -- & $\star$
 \\ \hline 
K & 39 & 19 & 20 & 3/2 & Y & Y & -- & -- & $\star$ &
O & 17 & 8 & 9 & 5/2 & -- & -- & Y & Y & $\star\star$
 \\ \hline 
V & 51 & 23 & 28 & 7/2 & Y & Y & -- & -- & $\star$ &
S  & 33 & 16 & 17 & 3/2 & -- & -- & Y & Y & 
 \\ \hline 
Co & 59 & 27 & 32 & 7/2 & Y & Y & -- & -- & &
Ca & 41 & 20 & 21 & 7/2 & -- & -- & Y & Y & $\star\star$
 \\ \hline 
Ga & 69 & 31 & 38 & 3/2 & Y & Y & -- & -- & &
Ca & 43 & 20 & 23 & 7/2 & -- & -- & Y & Y & $\star$
 \\ \hline 
Ga & 71 & 31 & 40 & 3/2 & Y & Y & -- & -- & &
Ni & 61 & 28 & 33 & 3/2 & -- & -- & Y & Y & $\star$
 \\ \hline 
Rb & 87 & 37 & 50 & 3/2 & Y & Y & -- & -- & $\star$ &
Ge & 73 & 32 & 41 & 9/2 & -- & -- & Y & Y &
 \\ \hline 
In & 113 & 49 & 64 & 9/2 & Y & Y & -- & -- & &
Sr & 87 & 38 & 49 & 9/2 & -- & -- & Y & Y &
 \\ \hline 
Sb & 121 & 51 & 70 & 5/2 & Y & Y & -- & -- & &
Zr & 91 & 40 & 51 & 5/2 & -- & -- & Y & Y & $\star$
 \\ \hline 
La & 139 & 57 & 82 & 7/2 & Y & Y & -- & -- & $\star$ &
Gd & 155 & 64 & 91 & 3/2 & -- & -- & Y & Y &
 \\ \hline 
Pr & 141 & 59 & 82 & 5/2 & Y & Y & -- & -- & $\star$ &
Gd & 157 & 64 & 93 & 3/2 & -- & -- & Y & Y &
 \\ \hline 
Re & 185 & 75 & 110 & 5/2 & Y & Y & -- & -- & &
Er & 167 & 68 & 99 & 7/2 & -- & -- & Y & Y &
 \\ \hline 
Re & 187 & 75 & 112 & 5/2 & Y & Y & -- & -- & &
Yb & 173 & 70 & 103 & 5/2 & -- & -- & Y & Y &
 \\ \hline 
Bi & 209 & 83 & 126 & 9/2 & Y & Y & -- & -- & $\star\star$ &
U & 235 & 92 & 143 & 7/2 & -- & -- & Y & Y &
 \\ \hline 
\end{tabular}

\end{center}
\narrowtext

Substances sensitive to proton parameters for Lorentz violation
are listed on the left-hand side of Table 3,
while those sensitive to neutron parameters 
are listed on the right-hand side.
The quantities
$A$, $Z$, $N$, and $I$ designate 
atomic weight, 
proton number, 
neutron number, 
and nuclear spin, 
respectively.
A symbol Y in a column labeled $D_w$ or $Q_w$
indicates sensitivity 
of the dipole or quadrupole energy shifts of the substance,
respectively,
to parameters for Lorentz violation for particles of type $w$.
In the left (right) half of this table,
all appearances of Y correspond to an odd proton (neutron) number,
and the neutron (proton) number is closed-shell 
\cite{fn4}.
Substances designated by the symbol $\star$
have magic neutron (proton) number,
while substances designated by the symbol $\star\star$
have both magic neutron (proton) number
and proton (neutron) number equal to a magic number plus one.
It seems plausible that these substances 
are most likely to have nuclear sensitivity to Lorentz violation 
depending only on a single valence proton (neutron).
The case of $^{91}$Zr is an exception,
in that the neutron number is a magic number plus one,
but the proton number is not magic.
Although it is \it not \rm a naturally abundant substance,
we have included $^{41}$Ca in the table because
it is relatively stable (lifetime $\simeq 10^5$ yr)
and it has magic proton number and neutron number equal to 
a magic number plus one.
We have also included the neutron itself in the table for completeness,
although technical challenges would need to be overcome
to perform Lorentz-violation experiments with (cold) neutrons. 

For future clock-comparison experiments,
the dual nuclear Zeeman $\he$-$^{129}$Xe maser already in operation
\cite{stoner}
could provide an interesting limit
on neutron parameters for Lorentz violation
because the $I=\half$ nucleus $^{129}$Xe is sensitive
to dipole energy shifts from neutron parameters
(in a complete nuclear model, it would probably also 
be sensitive to dipole energy shifts from proton parameters).
Within the Schmidt model,
the coefficients $\be_n$, $\ga_n$, $\de_n$, $\ka_n$, $\la_n$
for both $\he$ and $\xe$ are identical,
which would lead to a relatively clean bound.
Suppose an experiment 
with the quantization axis in the equatorial plane
produces a bound of $\ve_{1,J}$,
$J = X,Y$,
on sidereal variations 
of the free-running $\he$ frequency using $\xe$ as a reference.
Within the Schmidt model,
we find this would yield the bounds
\beq
|-3.5\tildeb_J^n+0.012\tilded_J^n
+0.012\tildeg_{D,J}^n | \lsim 2\pi\ve_{1,J} , 
\label{hexe}
\eeq
where the ratio of gyromagnetic ratios 
has been taken as $g^3/g^{129}\simeq 2.75$.
The factor of $-3.5$ is relatively large and 
compares favorably with the corresponding factor 
of $-2/3$ for the $\hg$-$\cs$ case,
so even a comparable precision for $\ve_{1,J}$
using the dual $\he$-$\xe$ maser
would represent an improved constraint on 
parameters for Lorentz violation 
by more than a factor of 5. 

Another interesting possibility would emerge from 
the development of a dual $\he$-$\ne$ maser
or a dual $\he$-$^{87}$Rb maser 
\cite{stoner2}. 
Table 3 shows that $\he$ is sensitive purely 
to dipole energy shifts from neutron parameters 
for Lorentz violation.
The $\he$ and quadrupole $\ne$ sensitivities
are discussed above and in Appendix \ref{hene}
in the context of the experiment of Chupp {\it et al.}
The dipole $\ne$ sensitivity within the Schmidt model includes 
$\tildeb^n_J$, $\tilded^n_J$, $\tildeg^n_{D,J}$,
and $\tildec^n_{Q,J}$,
though in a realistic nuclear model
$\ne$ would probably also be sensitive 
to parameters for Lorentz violation for the proton.
Table 3 also shows that $^{87}$Rb 
is a theoretically favorable substance.
A quadrupole measurement in a $\he$-$^{87}$Rb maser using the 
$\ket{\frac 32, \frac 32} \to \ket{\frac 32, \frac 12}$
or
$\ket{\frac 32, -\frac 12} \to \ket{\frac 32, -\frac 32}$
$^{87}$Rb transitions (but not both with equal weight)
therefore has the potential to provide an unusually clean bound 
on proton parameters for Lorentz violation.

\section{Summary}

In this work,
we have analyzed clock-comparison experiments
in the context of a general extension of the 
standard model and quantum electrodynamics
allowing for Lorentz and CPT violation.
In this theory,
both dipole and quadrupole shifts of atomic or ionic energy levels
are predicted and would produce sidereal and semi-sidereal 
time dependences of the signal.
We have obtained explicit formulae for these effects
that can be applied to existing and future experiments
and have demonstrated that the experimental results
already available place interesting constraints on 
certain combinations of the parameters in the theory.

Our expressions show that experiments performed 
with different atoms or ions typically test inequivalent quantities
as a result of possible variations of the parameters for Lorentz violation
with the species of elementary particle.
Indeed,
no two of the experimental bounds obtained to date 
involve identical linear combinations of parameters,
and the sensitivities of the two most recent experiments 
have no overlap at all.

The variety of high-precision experiments already performed
allows a region of the parameter space to be excluded.
However,
the exact specification of this region is theoretically uncertain 
because for the most part the bounds are obtained 
from atoms or ions with relatively involved nuclear structure. 
Some regions of the attainable parameter space 
are as yet unconstrained by clock-comparison experiments.

We have considered the issues involved in producing 
theoretically favorable bounds,
and have listed some naturally abundant substances 
that may be of potential interest for future tests.
The exceptional degree of precision attainable
offers potential sensitivity to Lorentz-violating effects 
from the Planck scale and 
ensures that future clock-comparison experiments 
remain among the most attractive possibilities
for detection of any nonzero effect that might exist in nature.

\acknowledgments
We thank R.\ Bluhm, L.R.\ Hunter, 
J.D.\ Prestage, R.E.\ Stoner, R.\ Walsworth,
and D.J.\ Wineland for useful discussions.
This work was supported in part
by the United States Department of Energy 
under grant number DE-FG02-91ER40661.

\begin{appendix}
\label{appendix1}

\section{Specific Experiments}
\label{bounds}

This Appendix contains remarks specific
to the experiments discussed in section \ref{application}.
Some issues relevant to the calculations leading to
Eq.\ \rf{generalbound} and Table 1 are presented.
Each experiment is considered under a separate heading.

All the experiments we consider apply a constant magnetic field 
of magnitude $B$ that fixes the quantization axis
of the atom or ion $W$.
Following the discussion in subsection \ref{geometry},
we define the $z$ axis to be aligned with this field.
Let $\vec{I}$, $\vec{J}$, and $\vec{F}$ represent the 
nuclear, electronic, and total angular momentum of $W$, 
respectively.
Where relevant,
we denote the corresponding quantum numbers by $I$, $J$, and $F$.
The degree to which $W$ is in an eigenstate of 
these operators is governed by a parameter
$\ze \approx { (g_J-g_I) \mu_B B }/{ E_{\rm hfs} }$,
where $g_J$ is the Land\'e $g$-factor of the electron cloud,
$g_I$ is the Land\'e $g$-factor of the nucleus,
$\mu_B$ is the Bohr magneton,
and $E_{\rm hfs}$ is the hyperfine splitting of the atom
\cite{breitrabi}.
For $J\not= 0$, $|g_I|\ll |g_J|$.
In most experiments that we consider, 
the applied magnetic field is small compared to
the internal interactions of $W$.
In this case,
$|\ze |\ll 1$,
$W$ is approximately in an eigenstate of 
$\vec{F}^2$ and $F_z$
with quantum numbers $F$ and $m_F$,
and the error introduced by approximating wave functions as 
eigenfunctions of $\vec{F}^2$ is suppressed by $\ze^2 \sim 10^{-12}$.
However,
in the experiment of Prestage {\it et al.} 
a relatively large magnetic field is applied to the $\ber$ ion.
In this case,
$|\ze |\gg 1$,
the ion is approximately in an eigenstate of
$I_z$ and $J_z$ with quantum numbers $m_I$ and $m_J$,
and the error due to approximating the $\ber$ wave function
to be an eigenfuction of $J_z$ and $I_z$ is suppressed by
${1}/{\ze^2} \sim 10^{-5} $.

\subsection{$^9$Be$^+$ and H Maser}
\label{ber}

The experiment of
Prestage {\it et al.} \cite{prestage}
measures the frequency $\nu$ of a $\ber$ transition
in a large (0.8194 T) magnetic field 
relative to the frequency of a hydrogen maser.
The $\ber$ transition is
$\ket{m_I, m_J} = \ket{-\frac 3 2,+\frac 1 2}
\to\ket{-\frac 1 2, +\frac 1 2}$.
The $H$-maser transition is 
$\ket{F, m_F} = \ket{1, 0} \to\ket{0, 0}$.
The experiment searches for a time variation 
in the frequency $\nu$ of the form 
$\nu= \nu_0 + A_k P_k(\cos{\be(t)})$,
where $A_1$, $A_2$, and $A_3$ are constants,
$P_k$ denotes the $k$th Legendre polynomial,
and $\be(t)$ is the angle between the quantization
axis and a direction of spatial anisotropy.
The limits obtained on the three quantities $|A_k|$ are 
approximately 100 $\mu$Hz.

Within the theoretical framework of the standard-model extension,
the standard hydrogen-maser frequency 
is unaffected by Lorentz violation
\cite{bkr2}.
The sensitivity to Lorentz violations
therefore resides entirely in the $\ber$ ion.
This ion has a nucleus with $I=3/2$ 
surrounded by an electron cloud with $J=1/2$,
so the nucleus could be sensitive 
in principle to dipole, quadrupole,
and octupole energy shifts,
while the electron cloud could be sensitive 
to dipole energy shifts.
However,
the transition frequency in the experiment is 
effectively insensitive to electron parameters for Lorentz violation
because $\De m_J\simeq 0$.
The formulae of subsection \ref{geometry} therefore apply
with $F$ replaced by the nuclear spin $I$.

The theoretical time variation $\nu(t)$ of the frequency $\nu$ 
can be obtained by applying Eq.\ \rf{tdatoms}
to the two energy levels involved.
In the experiment,
the magnetic field is at an angle of $\ch\simeq 118^{\circ}$ 
with respect to the Earth's rotation axis.
The various constants defined in Eq.\ \rf{bgdkl}
can be calculated approximately with the methods of
subsection \ref{coexva}.
The $\ber$ nucleus consists of 4 protons and 5 neutrons.
The Schmidt model predicts that a single neutron 
in a $p_{3/2}$ state
carries the entire nuclear angular momentum,
in agreement with the shell-model prediction 
that each valence nucleon is in a $p_{3/2}$ state.
The resulting values of the constants 
are given in Table 1.

The theoretical expression for $\nu(t)$ 
can be compared to the experimental fit for $\nu$.
This gives bounds of the form in Eq.\ \rf{generalbound},
where the constants are specified in Table 1.

\subsection{$^{201}$Hg and $^{199}$Hg}
\label{hghgg}

The experiment of Lamoreaux {\it et al.} \cite{lamoreaux}
compares precession frequencies 
of $\hgg$ and $\hg$ atoms in a weak magnetic field.
The electron clouds of both types of atom have $J=0$
in the ground state,
so the corresponding atomic states 
can be labeled $\ket{I, m_I}$.
The precession frequencies
arise from $\De m_I=1$ transitions.
The experiment searches for 
possible sidereal or semi-sidereal time variations 
in the frequency difference,
yielding an upper bound of about half a microhertz.

The $\hg$ nucleus has $I=1/2$
and is sensitive only to dipole shifts,
while the $\hgg$ nucleus has $I=3/2$
and is sensitive to dipole, quadrupole,
and octupole shifts.
The formulae of subsection \ref{geometry} apply
with $F$ replaced by the nuclear spin $I$.
The possible time variations in the 
observed frequency difference
can be found within the present framework
by using Eq.\ \rf{tdatoms} for each of the
energy levels involved in the transitions.
The magnetic field in the experiment lies in the 
Earth's equatorial plane,
so $\ch=\pi/2$
and many of the geometrical factors 
described in subsection \ref{geometry}
simplify.

The $\hgg$ nucleus has 80 protons and 121 neutrons,
while the $\hg$ nucleus has 80 protons and 119 neutrons.
The nuclear shell model predicts
that the ground-state properties of $\hgg$ and $\hg$ are
determined by the neutrons.
This implies the vanishing of all 
coefficients of the form \rf{bgdkl} for the proton
and would mean that both isotopes are sensitive only
to neutron parameters for Lorentz violation.
Both isotopes have valence protons and neutrons,
however,
so it is likely that a more realistic model
would produce nonzero coefficients \rf{bgdkl}
for protons too
and therefore that both nuclei are sensitive 
to proton and neutron parameters for Lorentz violation.

The Schmidt model indicates that
the angular momentum of the $\hg$ nucleus 
is carried by a single neutron in a $p_{1/2}$ state.
Naively,
this is at odds with the shell model, 
which implies each valence neutron is in an $i_{13/2}$ state.
However,
when there are nearly degenerate states 
with different orbital angular momenta $l$,
the shell model also suggests that
protons or neutrons prefer to pair in states of high $l$. 
This would mean that the $i_{13/2}$ shell is closed 
preferentially to shells immediately below it with lower $l$.
The shells immediately below $i_{13/2}$ are
$p_{1/2}$ and $p_{3/2}$,
so the Schmidt-model prediction is compatible 
with that from the shell model. 
A similar discussion applies to the $\hgg$ nucleus,
except that the single neutron is in a $p_{3/2}$ state.

Calculating the coefficients in Eq.\ \rf{bgdkl}
according to the methods of subsection \ref{coexva}
yields the results given in Table 1.
In converting the actual experimental bounds to the form
of Eq.\ \rf{generalbound} with the constants given in Table 1,
we have for simplicity approximated 
the $\hgg$ precession frequency as involving only the transition
$\ket{\frac 32, \frac 32} \to \ket{\frac 32, \frac 12}$.
A more accurate expression involving also the transition 
$\ket{\frac 32, -\frac 12} \to \ket{\frac 32, -\frac 32}$
could be obtained following the detailed analysis 
in Ref.\ \cite{lamoreaux},
but the results remain essentially unchanged.
Note that the nonzero value of $v$ in Table 1 reflects 
the ratio of gyromagnetic ratios of the two Hg isotopes,
$g_{201}/g_{199}\simeq -0.37$,
and the corresponding dependence 
of the sidereal bounds on both $\hgg$ and $\hg$.
In contrast,
the semi-sidereal bounds depend only on $\hgg$,
in accordance with its sensitivity to quadrupole shifts.

\subsection{ $^{21}$Ne and $^{3}$He }
\label{hene}

The experiment of Chupp {\it et al.} \cite{chupp}
searches for quadrupole shifts in $\ne$ precession frequencies 
relative to a reference precession frequency in $\he$,
placing a bound of about half a microhertz on possible
semi-sidereal variations.
The electron clouds of $\ne$ and $\he$ both have $J=0$
in the ground state,
so in a weak magnetic field 
only the nuclear angular momenta are relevant
and the corresponding atomic state 
can be labeled $\ket{I, m_I}$.
The experiment is insensitive
to electron parameters for Lorentz violation,
and the formulae of subsection \ref{geometry} apply
with $F$ replaced by the nuclear spin $I$.
The $\he$ nucleus has $I=1/2$
and is therefore sensitive only to dipole shifts,
while the $\ne$ nucleus has $I=3/2$
with sensitivity in principle to dipole, quadrupole,
and octupole shifts.

The shift in each energy level is given by 
Eq.\ \rf{tdatoms}
and can be used to deduce the 
possible time variations of the signal frequency 
in the present theoretical framework.
The magnetic field in the experiment is
perpendicular to the Earth's rotation axis,
so $\ch=\pi/2$
and many geometrical factors in subsection \ref{geometry} vanish.
Since the experiment bounds only semi-sidereal frequencies,
which are independent of dipole energy shifts,
the possible dipole energy shifts in both $\ne$ and $\he$ 
have no effect on the experiment.

The $\ne$ nucleus consists of 10 protons and 11 neutrons.
According to the shell model,
the ground-state properties of $\ne$ depend only
on the neutrons,
which suggests all coefficients of the type \rf{bgdkl}
for the proton must vanish and 
would imply the experiment is insensitive to
proton parameters for Lorentz violation.
However,
neither the protons nor the neutrons lie in a closed nuclear shell,
so it is likely that in reality the experiment
does have sensitivity to proton parameters for Lorentz violation.

In the Schmidt model,
the ground-state properties of 
$\ne$ and $\he$ are determined by a single neutron 
in a $p_{3/2}$ and an $s_{1/2}$ state, respectively.
This assignment for $\ne$ 
would appear to contradict the shell-model prediction
that each valence neutron is in a $d_{5/2}$ state.
It is,
however, 
plausible within the shell model that the $d_{5/2}$ shell is closed
preferentially to the states immediately below it in energy,
namely $p_{1/2}$ and $p_{3/2}$.
This argument for $\ne$ is weaker than 
the corresponding argument for $\hg$ 
in subsection \ref{hghgg} because the $p_{1/2}$ shell
relevant for $\ne$ is not merely closed 
but corresponds also to a magic number.
Since in any event a complete shell-model calculation
would still be inadequate in that the dependence
on proton parameters for Lorentz violation
would be missing,
we present only the Schmidt-model values in this work. 

The results of the calculation produce bounds
of the form of the last two equations in Eq.\ \rf{generalbound},
with coefficients given in Table 1.
To match the actual experimental bounds to this form,
we have for simplicity approximated 
the $\ne$ precession frequency as involving only the transition
$\ket{\frac 32, \frac 32} \to \ket{\frac 32, \frac 12}$.
A more accurate expression involving also the transition 
$\ket{\frac 32, -\frac 12} \to \ket{\frac 32, -\frac 32}$
could be obtained with the methods of Ref.\ \cite{chupp},
but this has no substantial effect on the results.

\subsection{ $^{199}$Hg and $^{133}$Cs }
\label{hgcs}

The experiment of Berglund {\it et al.} \cite{berglund}
bounds the possible sidereal time dependence of 
$\hg$ and $\cs$ 
precession frequencies.
The procedure uses a weak magnetic field to split
the ground states of the $\hg$ and $\cs$ atoms.
Denote the associated frequencies by  
$\nu^{133}$ and $\nu^{199}$.
The experiment measures the difference $\De B$
between the effective magnetic fields measured by the
$\hg$ and $\cs$ atoms.
This can be written
$\De B \equiv \nu^{199}/g^{199} - \nu^{133}/g^{133}$,
where 
$g^{199}\simeq 0.759$ kHz/G 
and 
$g^{133}\simeq 350$ kHz/G
are the gyromagnetic ratios of 
$\hg$
and 
$\cs$, 
respectively.
We take the experimental bound obtained
as a limit on possible sidereal variations of the frequency
difference $g^{199}\De B$ at the level of about 100 nHz.

The electron cloud of the $\hg$ atom 
in its ground state has $J=0$
and its nucleus has $I=1/2$,
so it is sensitive only to dipole shifts
and is insensitive to electron parameters for Lorentz violation.
See subsection \ref{hghgg} for more information about $\hg$.
In contrast,
the ground state of the $\cs$ atom 
has an electron cloud with 55 electrons in a $J=1/2$ state 
and a nucleus with $I=7/2$. 
The $\cs$ states relevant to the experiment 
have total angular momentum $F=4$,
so in principle sensitivity to nonzero energy shifts
of multipolarity up to order 8 would be possible.
However,
in the present framework the sidereal frequency dependences
bounded by the experiment can depend only on
dipole and quadrupole energy-level shifts.
The relevant shifts leading to 
possible time variations in the signal
are given by Eq.\ \rf{tdatoms}.
In the experiment,
the quantization axis is always perpendicular to 
the Earth's rotation axis
so $\ch=\pi/2$,
which simplifies the formulae in subsection \ref{geometry}.

The outer electronic shell of $\cs$ consists of a single
valence electron in a 6s state.
Since the closed shells do not contribute to dipole 
or quadrupole energy shifts,
only the valence electron is relevant.
It is straightforward to calculate the 
contributions to the coefficients in Eq.\ \rf{bgdkl}
for the electron,
using the expressions given in subsection \ref{coexva}.

The $\cs$ nucleus contains 55 protons and 78 neutrons.
The shell model suggests that the $\cs$
ground-state properties are independent of the neutrons. 
Since the $\hg$ properties do depend on neutrons,
even in the shell-model approximation
the experimental results are sensitive to
contributions from all three species of particle.
Moreover,
since neither the protons nor the neutrons lie in a 
closed nuclear shell,
the $\cs$ atom alone is likely to be sensitive
to parameters for Lorentz violation from all three species.
For simplicity and definiteness,
we limit the analysis in this paper to the Schmidt model,
for which the only significant nucleon is a proton
in a $g_{7/2}$ state
(in agreement with the shell model).

In the context of the present framework,
the bounds obtained in the experiment
take the form of the first two equations
in Eq.\ \rf{generalbound}.
The values of the coefficients are given in Table 1,
where the transitions have been taken as
$\ket{I,m_I}= \ket{\frac 1 2 , +\frac 1 2}\to 
\ket{\frac 1 2 ,-\frac 1 2}$ in $\hg$
and $\ket{F,m_F}=\ket{4,4}\to \ket{4,3}$ in $\cs$.
Note that the parameter $v$ is small,
primarily because the ratio of gyromagetic ratios
$|g^{199}/g^{133}|$ is small.

\end{appendix}

\end{multicols}

\begin{references}
\bibitem{fn1}
The term `clock-comparison experiment' is used throughout
this work to denote a generic class of experiments 
comparing transition frequencies in atoms or ions,
whether the experiment uses actual atomic clocks,
sensitive magnetometers, or other techniques.
\bibitem{fn2}
The clock-comparison experiments 
of Refs.\ \cite{hughes,drever,prestage,lamoreaux,chupp,berglund}
are distributed over several decades,
so in principle a comparison of their bounds 
should allow for this precession.
We neglect this effect here.
\bibitem{fn4}
In a nuclear shell model,
the magic numbers are 2, 8, 20, 28, 50, 82, 126, and 184 \cite{mayer}.
A closed-shell number is one corresponding to a closed nuclear shell,
in which the protons (neutrons) occupy a state with zero orbital
and spin angular momenta: 2, 6, 8, 14, 16, 20, 28, $\ldots$.
\end{references}
\end{document}